\documentclass[preprintnumbers,floats,prd,aps,nofootinbib]{revtex4}
\usepackage{amssymb,amsmath,graphicx}

\def\la{~\mbox{\raisebox{-.6ex}{$\stackrel{<}{\sim}$}}~}
\def\ga{~\mbox{\raisebox{-.6ex}{$\stackrel{>}{\sim}$}}~}

\begin{document}
\title{
Exact identification of the radion and its coupling
to the observable sector
}

\author{Lev Kofman, Johannes Martin and Marco Peloso}
\affiliation{
  CITA, University of Toronto\\
  Toronto, ON, Canada, M5S 3H8
}
\date{\today}

\begin{abstract}

Braneworld models in extra dimensions can be tested in laboratory by
the coupling of the radion to the Standard Model fields. The
identification of the radion as a canonically normalized field
involves a careful General Relativity treatment: if a bulk scalar is
responsible for the stabilization of the system, its fluctuations are
entangled with the perturbations of the metric and they also have to
be taken into account (similarly to the well-developed theory of
scalar metric perturbations in $4$D cosmology with a scalar
field). Extracting a proper dynamical variable in a warped
geometry/scalar setting is a nontrivial task, performed so far only in
the limit of negligible backreaction of the scalar field on the
background geometry. We perform the general calculation, diagonalizing
the action up to second order in the perturbations and identifying the
physical eigenmodes of the system for any amplitude of the bulk
scalar. This computation allows us to derive a very simple expression
for the exact coupling of the eigenmodes to the Standard Model fields
on the brane, valid for an arbitrary background configuration. As an
application, we discuss the Goldberger-Wise mechanism for the
stabilization of the radion in the Randall-Sundrum type models. The
existing studies, limited to small amplitude of the bulk scalar field,
are characterized by a radion mass which is significantly below the
physical scale at the observable brane.  We extend them beyond the
small backreaction regime. For intermediate amplitudes, the radion
mass approaches the electroweak scale, while its coupling to the
observable brane remains nearly constant. At very high amplitudes, the
radion mass instead decreases, while the coupling sharply
increases. Severe experimental constraints are expected in this
regime.

\end{abstract}

\preprint{CITA-2004-02}
\maketitle

\section{Introduction}

Braneworlds embedded in higher dimensions have opened new interesting
directions in high energy particle physics, general relativity, and
cosmology. The Randall-Sundrum (RS) model with warped $5$D bulk geometry
between two plane-parallel orbifold branes~\cite{RS1} has been
particularly studied. Its low energy spectrum contains a scalar field,
denoted as radion $Q$, which is associated to the inter-brane
modulus. The radion is coupled to Standard Model (SM) fields living on
one brane, which gives very interesting phenomenology for accelerator
experiments~\cite{phenomenology}. The coupling is of the form
\begin{equation}
\label{int1}
{\cal L}_{int}=\frac{Q}{\Lambda} \, T^{\mu}_{\mu} \,,
\end{equation}
where $T^{\mu}_{\mu}$ is the trace of the energy momentum of the SM fields,
$Q$ refers to the radion field, while the coupling $\Lambda$ is close
to the electroweak scale~\cite{GW,GW2,csaki1,csaki2}.

The radion originates from the scalar fluctuations of the bulk
geometry. In the RS model~\cite{RS1} the two branes are at indifferent
equilibrium: the interbrane distance is a free parameter,
corresponding to a massless radion. The situation is changed if a bulk
scalar field $\phi$ is introduced to stabilize the system, as in the
Goldberger-Wise mechanism~\cite{GW}.  The identification and the study
of the properties of the radion field are less straightforward in this
case. The initial studies, including the original work~\cite{GW}, were
based on the effective $4$D theory, obtained integrating an
approximate ansatz for the geometry and the scalar field profile along
the extra coordinate. However, the introduction of the bulk scalar
makes the gravity/scalar system more complex, and requires a
self-consistent treatment based on the $5$D Einstein
eqs.~\cite{Dewolfe,GR}. In particular, the scalar fluctuations $\Phi$
of the bulk metric are sourced by the fluctuations $\delta \phi$ of
the bulk scalar field, and the coupled linearized system $\left( \Phi,
\delta \phi \right)$ also has to be treated self-consistently. The $5$
dimensional equations for the system were derived in the basic
papers~\cite{Tanaka:2000er,csaki2}. The analysis of \cite{csaki2} is
particularly advanced and clear with respect to the coupling of the
modes, so we will often refer to it for comparison. The properties of
the radion obtained in these studies are nonetheless in good agreement
with the ones from the $4$D effective theory (the agreement is
actually much better than what has been claimed sometimes in the
literature, as we will discuss below).

All the existing studies share one limitation: it is always assumed
that the bulk scalar field has a negligible backreaction on the
background geometry (the only effect being that it lifts up the flat
direction corresponding to the radion). Phenomenological studies have
so far been restricted to this case, first because it is only in this
limit that the bulk geometry is (approximately) AdS as in the RS
proposal, but also because it is the only case for which the
properties of the radion were known. Indeed, in this limit the
contribution of $\delta \phi$ to the wave function of the radion is
negligible, and the radion behaves approximately like the metric
perturbation $\Phi\,$~\cite{csaki2}. As a consequence, the exact
diagonalization of the coupled $\left( \Phi, \delta \phi \right)$
system in not required in this limit.

In contrast to these works, the aim of the present paper is to study the
scalar perturbations for the general situation in which the bulk
scalar $\phi$ has arbitrary amplitude. The main motivation is that
very small $\phi$ should not be considered as completely natural,
since the idea of RS is to avoid strong hierarchies besides the one
given by the warp factor.~\footnote{A larger $\phi$ is also more
  natural (at least for the Goldberger-Wise mechanism) since it
  increases the fraction of initial conditions for which the
  stabilization is achieved, if the radion is initially displaced from
  the stabilized value. See Appendix~\ref{ap3} for a discussion.} There is
no incompatibility between a sizable $\phi$ and the main idea
of~\cite{RS1}: although the bulk geometry will be in general different
from pure AdS, one can still naturally achieve a strong warping at the
observable brane, as needed to solve the hierarchy problem. In particular,
this is also true for the Goldberger-Wise mechanism \cite{GW} and its
generalizations. The main obstacle is that the mixing between the
different scalar perturbations cannot any longer be neglected, and the
coupled system of perturbations has to be diagonalized.  This study is
still missing in the existing literature, and it is the main result of the
present work. As we will show, the coupled $\left( \Phi, \delta \phi
\right)$ pair contains a single dynamical variable which we denote by $v$,
and which is a linear superposition of the pair. The equation for $v$
reduces to an oscillator-like equation. More importantly, the action of
the system can be shown to be diagonal in terms of the $4$D Kaluza-Klein
modes of $v\,$, which we denote by $Q_n\,$. In this way, we obtain the
{\it exact} $4$D action for the physical scalar modes of the theory. The
action differs from the effective (approximate) $4$D theory which has been
computed in the literature by neglecting the $\delta \phi$ modes, and -
moreover in general - by starting from a $5$D ansatz which {\it does not}
satisfy the linearized equations for the perturbations (the effective
action is derived and discussed in Appendix~\ref{ap3}, to emphasize the
difference from the exact $4$D theory discussed in the rest of the paper).

The final result can be easily summarized. Decomposing the metric
perturbation in terms of the $4$D KK modes, $\Phi = \sum_n {\tilde
\Phi_n} \left( y \right) \, Q_n \left( x \right)$ ($y$ denoting the
extra coordinate), we find that the quadratic action for the scalar
modes can be cast in the diagonal form
\begin{equation}
S = \sum_n S_n = \sum_n C_n \int d^4 x \; Q_n \left[ \, \Box - m_n^2 \,
\right] Q_n \,\,,
\end{equation}
where $m_n^2$ are the physical masses obtained from an eigenvalue
boundary problem in the extra dimension (see below). As we remarked,
the radion corresponds to the mode with the lowest mass. The fields $Q_n$
are scalars fields of the exact $4$D theory, and their decoupled
actions can be trivially quantized. The coefficients $C_n$ are given
by
\begin{equation}
C_n = \frac{3 \, M^3}{2} \int_0^{y_0} \frac{d y}{A} \, \left[ \frac{3 \,
M^3}{2 \, \phi'^{\,2}} \, \left( A^2 \, {\tilde \Phi}_n \right)'^{\,2} \, 
 + \left( A^2 \, {\tilde \Phi}_n \right)^2 \right] \,\,.
\label{result}
\end{equation}
In this expression, $A$ is the warp factor, $M$ the fundamental
scale of gravity in $5$ dimensions, and prime denotes differentiation
with respect to $y\,$. This result is particularly simple and
completely general, since it holds for arbitrary background
configurations $A$ and $\phi\,$. The derivation, to be discussed in
detail below, exploits the strong similarity of the system with the
one of scalar metric fluctuations in the scalar field inflationary
cosmology. This analogy allows us to identify the dynamical variable $v$,
which in inflationary cosmology is known as the Mukhanov-Sasaki variable
\cite{mukhanov,sasaki}, and which is the starting point of our 
computation.

The requirement that the modes $Q_n$ are canonically normalized, $C_n
\equiv 1/2\,$, fixes the normalization of ${\tilde \Phi}_n\,$, which
in turn defines the strength of the coupling of the radion and of the
other KK modes to the SM fields on the observable brane (the
computation and diagonalization of the action is indeed necessary for
this task. The normalization of the modes cannot be derived from the
Einstein equations for the perturbations, which are all linear in the
perturbations).  Similarly to the initial expression~(\ref{int1}), the
coupling $\Lambda_n$ of the radion and the other KK modes to brane
fields is given by
\begin{equation}
{\cal L}_{{\rm int}, n} = \frac{Q_n}{\Lambda_n} \, T^\mu_\mu \quad \quad ,
\quad \quad \Big\vert \frac{1}{\Lambda_n} \Big\vert = \Big\vert
\frac{{\tilde \Phi}_n \left( y_0 \right)}{2} \Big\vert \,\,.
\label{int2}
\end{equation}
where $y_0$ is the location of the observable brane. Due to the
explicit knowledge of~(\ref{result}), we can determine the coupling of
all the perturbations for arbitrary background configuration. Our
general calculation confirms the accuracy of the results
of~\cite{csaki2} in the limit of small $\phi\,$.~\footnote{We thank
C.~Csaki for discussions and clarifications on this issue.} Our
results extend these calculations to a general amplitude of the field
$\phi$ and to an arbitrary geometry $A\,$.

The paper is organized as follows. In section~\ref{setup} we present the
action and the relevant equations of the system. The main result
summarized above is derived in section~\ref{quanti}, while it is applied
to a couple of specific examples in section~\ref{coupl}, one of which -
first proposed in~\cite{Dewolfe} - is a generalization of the
Goldberger-Wise mechanism~\cite{GW} for the stabilization of the radion of
the Randall-Sundrum model~\cite{RS1}. The existing studies of the
phenomenology of this model, limited to small amplitude of the bulk scalar
field, are characterized by a radion mass which is significantly below the
physical scale at the observable brane. This hierarchy decreases as $\phi$
increases, and, fore sizable values of $\phi\,$, the radion mass
approaches the electroweak scale. On the contrary, the coupling of the
radion to the observable sector remains constant for a wide range of
$\phi\,$. The radion mass reaches a maximal value for a finite value of
$\phi\,$. At higher amplitudes the mass decreases, while its coupling to
the SM sharply increases. In section~\ref{branecode} we verify numerically
some of these analytical results: we follow the oscillations of a
two-brane system around a static configuration with the numerical code
developed in~\cite{bc}; the frequency spectrum of the oscillations
exhibits resonances precisely in correspondence to the masses of the modes
given by the analytical computation. The higher resonances are excited by
the their coupling to the radion in the nonlinear regime. Once their
amplitude becomes small, they become constant in time, confirming the fact
that they are decoupled at the linear level. The results of the paper are
discussed in section~\ref{discussion}, where we also comment on more
general related issues. For example, we discuss preheating effects due to
the oscillations of the radion, and the linearized $4$D gravity on the
branes.  The paper is concluded by three Appendices. In the first one we
compute the quadratic action for the perturbations. Appendix~\ref{ap2}
presents analytical results in the small $\phi$ limit. Finally,
Appendix~\ref{ap3} contains the computation of the effective $4$D action
for the radion, and the comparison with the exact results discussed in the
rest of the paper.

\section{Set-up} \label{setup}

We start from the action
\begin{equation}
S = 2\int_0^{y_0} d^5 x \sqrt{- g} \left[ \frac{M^3}{2} R -
\frac{1}{2} \left( \partial \, \varphi \right)^2 - V \left( \varphi
\right) \right] 
- \sum_{i=1,2} \int d^4 x \sqrt{-\gamma} \left\{ 2M^3
\left[ K \right] + U \left( \varphi \right) \right\} \,\,,
\label{ac1}
\end{equation}
where the first term is the bulk action, and the sum contains
contributions from the two branes, located at $y=0$ and $y=y_0 \equiv
- y_0$ along the extra dimension. The overall factor of $2$
multiplying the first term accounts for the integration over the
intervals $[-y_0,0]$ and $[0,y_0]$, which are identified by an
$S^1/{\Bbb Z}_2$ symmetry. To shorten the notation, the bulk
cosmological constant is included in the bulk potential $V\,$, while
brane tensions are included in the brane potentials $U\,$. The
quantity $\left[ K \right]$ denotes the jump of the extrinsic
curvature at the two branes, while $M$ is the fundamental scale of
gravity in $5$ dimensions. We choose the following line element
\begin{equation}
d s^2 = A \left( y \right)^2 \left[ \left( 1 + 2 \Psi \right)
\eta_{\mu \nu} \, d x^\mu \, d x^\nu + \left( 1 + 2 \Phi \right) d y^2
\right] ,
\label{line}
\end{equation}
where Greek indices span the $1+3$ coordinates parallel to the flat
branes, and $\eta = {\rm diag } \left( -1, 1, 1, 1 \right) \,$.
Namely, we have conformal coordinates for the warped background
geometry, while the scalar metric perturbations $\Psi$ and $\Phi$ are
expressed in the (generalized) longitudinal gauge.~\footnote{We need
to consider a smaller set of scalar metric perturbations with respect
to the one discussed in brane cosmology, when the induced metric on the
brane is of $FRW$ type, see for instance~\cite{bdbl}. When, as in our
case, the brane is maximally symmetric, these other modes are
decoupled from the ones we consider here, and they are actually
components of the massive gravitational waves~\cite{andrei}. Scalar
perturbations also include the displacements $\xi_i$ of the two branes
along the $y$ coordinate. However, in absence of fluids or kinetic
terms for the scalar on the brane, $\xi_i =0\,$~\cite{FK}, so that we
do not need to include them in the calculation. In Section~\ref{coupl}
we consider brane fields, to compute their interaction with the radion
and the KK modes. However, it is accurate to treat them as probe
fields (consistently with the interaction picture), which do not
modify the setting discussed here.} Perturbations of the bulk scalar
field are denoted $\varphi \left( x, y \right) = \phi \left( y \right)
+ \delta \phi \left( x, y \right)\,$. We are interested in static
warped configurations (and, in addition, flat branes), so that the two
background quantities $A$ and $\phi$ depend on the extra coordinate
$y$ only. Concerning the background quantities, variation of the
action~(\ref{ac1}) gives the bulk equations (here and in the
following, prime denotes derivative wrt $y\,$)
\begin{eqnarray}
&&\phi'' + 3 \, \frac{A'}{A} \, \phi' - A^2 \, V' = 0 \,\,, \nonumber\\
&& \frac{A''}{A} = 2 \frac{A'^2}{A^2} - \frac{\phi'^2}{3 \, M^3} \,\,,
\nonumber\\
&&6 M^3 \frac{A'^2}{A^2} = \frac{\phi'^2}{2} - A^2 \, V \,\,,
\label{back}
\end{eqnarray}
as well as the junction conditions
\begin{equation}
\frac{A'}{A^2} = \mp \frac{U}{6 \, M^3} \quad,\quad\quad \frac{\phi'}{A} =
\pm \frac{U'}{2} \,\,,
\label{jump}
\end{equation}
where the upper and lower sign refers to the brane at $y=0$ and
$y=y_0\,$, respectively (prime on the potentials $V$ and $U$ denotes
differentiation with respect to $\phi\,$).

As with the background, not all of the equations for the
perturbations are dynamical, reflecting the fact that not all the
modes given above are physical. The system of linearized Einstein
equations for the perturbations include the two constraint
equations
\begin{eqnarray}
&&\Phi - \frac{A}{A'} \Psi' - \frac{A \, \phi'}{3 \, M^3 \, A'} \,
\delta \phi = 0 \,\,, \nonumber\\
&&\Phi = - 2 \, \Psi \,\,.
\label{constraint}
\end{eqnarray}
as well as a dynamical equation. By using the constraint equations,
the latter can be written as
\begin{eqnarray}
\left[ \Box + \frac{d^2}{d \, y^2} - \frac{\left( 1/z
\right)''}{1/z} \right] {\cal U} = 0 \,\,,
\label{dyn1}
\end{eqnarray}
where we have defined
\begin{eqnarray}
z \equiv \sqrt{2}\frac{A^{5/2} \, \phi'}{A'} \quad \quad , \quad \quad
{\cal U} \equiv \frac{A^{3/2}}{\phi'} \, \Psi \,\,,
\label{defz}
\end{eqnarray}
and where $\Box$ is the D'Alambertian operator in $4$ d. Finally,
there are two junction conditions for the perturbations at the two
branes,
\begin{eqnarray}
&& \left( A^2 \, \Phi \right)' = \frac{2 \, A^2 \, \phi' \, \delta
\phi}{3 \, M^3} \,\,, \nonumber\\
&&\delta \phi' - \phi' \, \Phi = \pm \frac{1}{2} \, A \, U'' \,
\delta \phi \,\,.
\label{junpert}
\end{eqnarray}

The first of these conditions could have also be obtained from the two
constraint equations (\ref{constraint}) evaluated at the two branes.
The second condition instead contains additional, physically relevant
information, relating $\delta \phi$ to the ``mass parameters'' of the
scalar field at the two branes. As can be intuitively expected, very
high $U''$ force $\delta \phi = 0$ at the two branes. This is the case
which is typically considered in the literature (starting from the
original proposal~\cite{GW}), since the results simplify significantly
in this limit. In the next section we perform the identification and
quantization of the perturbations for arbitrary $U''$, and we give
general formulae valid for arbitrary bulk and brane potentials for
$\phi\,$. The examples of section~\ref{coupl} are instead restricted
to high $U''\,$.

\section{Quantization of the action for perturbations and the 
identification of the physical modes.} \label{quanti}

From the analysis of the previous section, we see that scalar metric
perturbations are coupled to the fluctuations of the bulk scalar field
responsible for the stabilization of the system. The covariant
formulation~(\ref{ac1}) fully underlines the symmetries of the
theory. However - as in all gauge theories - the price to pay is the
introduction of auxiliary degrees of freedom, so that it is not
straightforward to identify and decouple the physical modes of the
system. However, this is very important if we want to know their
properties, namely their masses and their coupling to the
(electroweak) sector on the observable brane. The mass spectrum for
the perturbations can be obtained directly from the equations of
motion. The easiest way to do so is to first identify some linear
combination of the perturbations for which the dynamical equation can
be cast in a Schr\"odinger-type
form~\cite{Tanaka:2000er,csaki2,FK,lorentzo}. This is the case for the variable
${\cal U}$ defined in eq.~(\ref{defz}), although other combinations
can be used (see below). After solving this equation, the spectrum of
the system can be obtained from the boundary
conditions~(\ref{junpert}) at the two branes.

This procedure, however, does not identify which combination $v$ of
the perturbations corresponds to the physical degrees of freedom, nor
does it determine the proper normalization of the modes, since the
bulk and boundary conditions are linear in the
perturbations.~\footnote{On a technical level, ${\cal U}$ is a
  combination ${\cal U} = c_1 {\cal U}_1 + c_2 {\cal U}_2$ of the two
  linearly independent solutions of equation~(\ref{dyn1}). The
  boundary conditions at the two branes allow the determination of the
  ratio $c_1/c_2\,$, as well as the physical mass of the mode (defined
  as the eigenvalue of $\Box\,$, see below), but they do not allow to
  determine $c_1$ and $c_2$ independently.} On the other hand, this
information is needed to determine the coupling of the physical modes
to the observable sector, see eq.~(\ref{int2}). To obtain it, one has
to diagonalize the action $S_2$ for these modes, which is computed by
expanding the initial action~(\ref{ac1}) up to second order in the
perturbations. This is a more difficult exercise than simply obtaining
the equations of motion for the perturbations, and indeed it is the
main original result of the present work.

We divide the discussion in two subsections. In the first one we
quantize the system for the case in which the bulk scalar field $\phi$
is absent. This is a particularly simplified situation, since in this
case there is only one physical scalar perturbation. This computation,
which is the only one given in the literature, allows us to determine
the normalization of the radion when the bulk scalar field is absent
or when its backreaction on the bulk geometry can be
neglected~\cite{csaki2}. As shown in~\cite{csaki2}, the contribution
of $\delta \phi$ to the wave function of the radion is negligible in
this limit, and the coupling of the radion to brane fields is
(approximately) given by the coupling of $\Phi\,$. In
subsection~\ref{sub2} we perform the general calculation, valid for a
bulk scalar field of any amplitude. This computation confirms the
accuracy of the analysis of~\cite{csaki2} in the limit in which the
backreaction of $\phi$ is small and can be treated perturbatively.  In
addition, it gives new interesting results. First, as we mentioned, it
provides the coupling of the radion to brane fields for arbitrary
values of $\phi\,$.  Second, it allows the computation of the coupling
of the other KK modes, which cannot be approximated by the analysis
of~\ref{sec:2A} even in the small $\phi$ limit. The reason for this is
the discontinuity in the number of modes with or without the bulk
scalar field: when $\phi$ is absent, there is only one physical scalar
degree of freedom (the radion). On the contrary, there is a whole
tower of KK modes when the scalar field is present (no matter how
small its amplitude is). Hence, the study of the action without $\phi$
does not provide any information about the KK modes.

\subsection{Computation without the bulk scalar field}
\label{sec:2A}

It is very instructive to study what happens when the scalar field is
absent. Expanding the action~(\ref{ac1}) at quadratic order in the
perturbations, and imposing $\Psi = - \Phi/2\,$, we find~\footnote{
The action~(\ref{anp}) is obtained by setting $\phi = \delta \phi = 0$
in the general expression~(\ref{ac2}) given below.}
\begin{eqnarray}
S_2 \left[ \varphi = 0 \right] = && 2M^3 \int d^5 x \, A^3
\left[  \frac{-3}{4} \eta^{\mu \nu} \partial_\mu \Phi \partial_\nu
\Phi + \frac{3}{2} \Phi'^2
+ 6 \frac{A'}{A} \Phi \Phi' + \left( 12 \frac{A'^2}{A^2}
- 3 \frac{A''}{A} \right) \Phi^2 \right] \,\,.
\label{anp}
\end{eqnarray}

From eq.~(\ref{constraint}) we have
\begin{equation}
\Phi' + 2 \frac{A'}{A} \Phi = 0 \;\;\; \Rightarrow \;\;\; \Phi \propto
A^{-2} \,\,.
\label{phinophi}
\end{equation}
Hence, in this case there is only one scalar eigenmode, corresponding
to the radion field, while the tower of KK modes is
absent~\cite{cgr}. We can then eliminate $\Phi'$ in favor of $\Phi\,$,
so that the action~(\ref{anp}) becomes
\begin{eqnarray}
&& S_2 \left[ \varphi = 0 \right] = 2 M^3 \int d^5 x A^3 \left[ -
\frac{3}{4} \eta^{\mu \nu} \partial_\mu \Phi \partial_\nu \Phi + 3
\left( 2 \, \frac{A'^2}{A^2} - \frac{A''}{A} \right) \Phi^2 \right]
\,\,.
\end{eqnarray}

The second term vanishes due to the background
equations~(\ref{back}). We thus recover the well known result of a
massless radion in the absence of any bulk field. Decomposing
\begin{equation}
\Phi \left( {\bf x}, y \right) = {\tilde \Phi} \left( y \right) Q
\left( {\bf x} \right) \,\,,
\end{equation}
the above action becomes
\begin{eqnarray}
S = - \frac{3 \, M^3}{2} \int_0^{y_0} d y \, A^3 \, {\tilde \Phi}^2
\int d^4 x \eta^{\mu \nu} \partial_\mu Q \partial_\nu Q \equiv -
\frac{1}{2} \int d^4 x \, \eta^{\mu \nu} \, \partial_\mu Q \,
\partial_\nu Q \,\,.
\end{eqnarray}
The last condition (i.e. the request that the $4$D field $Q$ is
canonically normalized) specifies the normalization of ${\tilde
\Phi}\,$, which in turns determines the coupling of the radion with
the fields on the observable brane. The quantization of the last
expression is straightforward.

When the bulk scalar is absent, the function $Q\,$, which encodes the
physical degree of freedom, obeys the free wave equation $\Box Q=0\,$.
The free bulk wave equation is satisfied by the bulk gravitational
waves $h_{AB}$. The $5$ bulk degrees of freedom of $h_{AB}$ are
projected to the $4$D brane as two tensor modes (corresponding to
usual $4$D gravitons), two gravi-vectors (which are however absent at
orbifold branes) and one gravi-scalar. When the bulk scalar is absent,
we suggest interpreting the massless radion $Q$ as the massless
gravi-scalar projection of the bulk graviton.

When the branes are stabilized by the bulk scalar field, the massless
gravi-scalar projection is absent~\cite{andrei}. However, this degree
of freedom does not disappear. The dynamical degree of freedom will be
in this case a linear combination of the metric perturbation $\Phi$
and the fluctuation of the scalar field $\delta \phi\,$, and it
emerges as a massive radion at the brane, rather than as a
gravi-scalar (see the next section for details). The situation is
similar to the re-shuffling of the degrees of freedom which occurs
during a symmetry breaking, where here the symmetry breaking is
associated with the stabilization of the interbrane distance.

\subsection{General computation with a bulk scalar field} \label{sub2}

The expansion of the action~(\ref{ac1}) at second order in the
perturbations gives
\begin{eqnarray}
&&S_2 =2 M^3 \!\! \int d^5 x \, A^3 \left\{ \eta^{\mu \nu} \!\! \left[
\frac{\partial_\mu \delta \phi \, \partial_\nu \delta \phi}{-\,2\ M^3}
+ 3 \partial_\mu \left( \Phi + \Psi \right) \partial_\nu \Psi \right]
- \frac{1}{2 \, M^3} \delta \phi'^2 + 6 \Psi'^2 + \frac{1}{M^3} \,
\phi' \delta \phi' \left( \Phi - 4 \Psi \right) + \right. \nonumber\\
&&\left. + 12 \frac{A'}{A} \Psi' \left( - \Phi + 2 \Psi \right) -
\left( \frac{\phi'^2}{M^3} - 12 \frac{A'^2}{A^2} \right) \left(
\frac{1}{2} \Phi^2 + 4 \Psi^2 \right) - \left[ \frac{1}{2 \, M^3} A^2
V^{''} \delta \phi^2 - \frac{1}{M^3} A^2 V' \delta \phi \left( \Phi +
4 \Psi \right) \right] \right\} + \nonumber\\
&&- 2\sum_i \int d^4 x A^4 \left[ 2 U \Psi^2 + 2 U' \delta \phi \Psi +
\frac{1}{4} U'' \delta \phi^2 \right] \,\,.
\label{ac2}
\end{eqnarray}
The derivation is given in Appendix~\ref{ap1}. Due to the presence of
the bulk scalar field and its fluctuations, the general
expression~(\ref{ac2}) is more involved than~(\ref{anp}), and the
quantization is more difficult. There are however similarities with
the well known problem of scalar perturbations in $4$D inflationary
cosmology with a scalar field, which fortunately can be exploited in
the case at hand.~\footnote{The reduction to the dynamical action that
we give here extends the one performed in~\cite{Mukhanov:jd,mfb} for
the case of cosmological perturbations. Although we show it here in
longitudinal gauge, we have actually derived it in arbitrary gauge for
the perturbations.} In the cosmological set-up, the inflaton field is
the analogous of the bulk scalar $\phi\,$, and the time $t$ is
analogous to the bulk coordinate $y\,$ (the scale factor $a \left( t
\right)$ is the analogue of the warp factor $A \left( y
\right)\,$). The evolution equations for the inflationary case
constitute an initial value problem (initial conditions specified at
some early time during inflation), which is in our case replaced by a
boundary value problem (junction conditions at the two branes). The
analogy cannot be pursued up to the quantization (in both cases, a
canonical quantization is performed, and the $t \leftrightarrow y$
analogy has to be abandoned). However, it allows us to identify the
dynamical variable $v$ which is the starting point for the
quantization of~(\ref{ac2}). There are two complications compared to
the inflationary case. The first is that, from the $4D$ point of view,
the system~(\ref{ac2}) contains an infinite KK tower of scalar
perturbations, while there is only one scalar mode in the inflationary
case (as it was also the case in the previous subsection). The second
is the presence of two branes as boundaries of the $y$
coordinate. Total derivatives in $y$ cannot be dropped (unlike the
cosmological situation, where the space has no boundaries). As we
shall see, they are actually crucial to ensure the orthogonality of
the different modes in the action.

Starting from the action~(\ref{ac2}), we first eliminate $\Phi$ by use
of the first of~(\ref{constraint}) (nondynamical equations, have to be
used for the quantization. They compensate for the presence of
non-physical degrees of freedom in the set of perturbations). We then
identify total derivatives, and simplify the final expression by use
of the background equations. Finally, we use $\Psi = - \Phi /2$ to
express $\Psi$ (and its derivatives) back in terms of
$\Phi\,$. Although the calculation is rather involved, the final
result is particularly simple
\begin{eqnarray}
S_2 &=& \frac{1}{2} \int d^5 x \, v \left[ \Box + \frac{d^2}{d y^2} -
\frac{z''}{z} \right] v + \nonumber\\
&&+ \int d^5 x \, \partial_y \, \left[ \frac{3 \, M^3 \, A^4}{4 \, A'}
\, \eta^{\mu \nu} \partial_\mu \Phi \partial_\nu \Phi - \frac{A^4}{A'}
\, \Phi \left( 2 \frac{A'}{A} \, \phi' - \frac{A^2 \, V'}{2} \pm
\frac{A \, U'' \phi'}{4} \right) \delta \phi \right] \,\,.
\label{ac3}
\end{eqnarray}
The variable $z$ was defined in~(\ref{defz}). The combination 
\begin{equation}
v \equiv z \left( - \frac{\Phi}{2} - \frac{A'}{A \, \phi'} \delta \phi
\right) = \frac{3 M^3}{z} \left( z \, {\cal U} \right)'
\label{defv}
\end{equation}
is the $5$D generalization of what is commonly denoted as Mukhanov-Sasaki
variable~\cite{mukhanov,sasaki}, and it is the dynamical variable in 
the bulk. It is gauge invariant, i.e. it does not change under
infinitesimal redefinitions of the coordinate system. The bulk
equation for $v$ is immediately obtained from the action~(\ref{ac3}),
and it can be shown to be equivalent to the dynamical eq.~(\ref{dyn1})
expressed in terms of ${\cal U}\,$
\begin{equation}
\left[ \Box + \frac{d^2}{d \, y^2} - \frac{z''}{z} \right] v = 0 \,\,.
\label{eqv}
\end{equation}

As we mentioned, the boundary terms cannot be neglected. Their
presence enforces the hermiticity of the Lagrangian operator.
Starting from only the bulk term, we get~\footnote{In the step denoted
  by dots we have differentiated the definition of $v\,$ and made use
  of~(\ref{junpert}) to eliminate $\Phi'$ and $\delta \phi'\,$.}
\begin{eqnarray}
&& \frac{1}{2} \int d^5 x \left\{ v_1 \left[ \Box + \frac{d^2}{d y^2}
- \frac{z''}{z} \right] v_2 - v_2 \left[ \Box + \frac{d^2}{d y^2} -
\frac{z''}{z} \right] v_1 \right\} = \frac{1}{2} \int d^5 x \left( v_1
\, v_2 ' - v_1' v_2 \right)' = \nonumber\\
&& = \dots = \int d^5 x \left[ \frac{A^4}{A'} \, \left( \Phi_1 \,
\delta \phi_2 - \delta \phi_1 \, \Phi_2 \right) \, \left( 2
\frac{A'}{A} \, \phi' - \frac{A^2 \, V'}{2} \pm \frac{A \, U''
\phi'}{4} \right) \right]' \,\,.
\end{eqnarray}
The last term precisely cancels with the difference of the boundary
terms in~(\ref{ac3}). This guarantees the orthogonality of the KK
modes of the scalar perturbations. The different modes arise from the
decomposition
\begin{equation}
F \left( x, y \right) = \sum_n {\tilde F}_n \left( y \right) \, Q_n
\left( x \right) \,\,,
\label{deco}
\end{equation}
where $F$ is any of $\delta \phi, \Phi,$ and $v\,$ (the same functions
$Q_n$ have to be used in order to satisfy the coupled
equations~(\ref{junpert}) and~(\ref{defv})). The equations for the
perturbation separate in $x$ and $y\,$. For example, eq.~(\ref{eqv})
enforces the two eigenvalue equations in the bulk
\begin{equation}
\left( \Box - m_n^2 \right) Q_n = 0 \quad\quad,\quad\quad
\left( \frac{d^2}{d y^2} - \frac{z''}{z} + m_n^2 \right) {\tilde v}_n
= 0 \,\,.
\end{equation}
Similarly, eq.~(\ref{dyn1}) gives an eigenvalue equation for ${\tilde
{\cal U}}_n\,$. Combining it with the constraint equations to
eliminate terms in ${\tilde \Phi}_n'$ and ${\tilde \Phi}_n''\,$, we
get
\begin{equation}
\left( 2 \frac{A'}{A} \phi' - \frac{A^2 V'}{2} \pm \frac{A U'' \phi'}{4}
\right) {\tilde \delta \phi}_n = - \frac{3 M^3}{4} m_n^2 {\tilde \Phi}_n
\,\,.
\end{equation}
Finally, we insert the decomposition~(\ref{deco}) into the
action~(\ref{ac1}). Making use of the last two equations, we get
\begin{eqnarray}
S  &=& \sum_{n,m} C_{n m} \int d^4 x \; Q_m \left[ \, \Box - m_n^2 \,
\right] Q_n \,\,, \nonumber\\
C_{n m} &\equiv& \frac{1}{2} \int_0^{y_0} d y \, {\tilde v}_m \,
{\tilde v}_n - \frac{3 M^3 A^4}{4 A'} {\tilde \Phi}_m {\tilde \Phi}_n
\Big\vert_0^{r_0} = \nonumber\\
&=& \frac{3 \, M^3}{2} \int_0^{y_0} \frac{d y}{A} \, \left[ \frac{3 \,
M^3}{2 \, \phi'^2} \, \left( A^2 \, {\tilde \Phi}_m \right)' \,
\left( A^2 \, {\tilde \Phi}_n \right)' + A^2 \, {\tilde \Phi}_m \, A^2
\, {\tilde \Phi}_n \right] \equiv C_n \, \delta_{m n} \,\,.
\label{ac4}
\end{eqnarray}
The second expression for $C_{m n}\,$ has been obtained relating $v$
and $\Phi$ through ${\cal U}$ (eqs.~(\ref{dyn1}) and~(\ref{defv})).
Hermiticity and orthogonality of different modes are manifest from
it.

This completes the diagonalization of the starting action in terms of
the different scalar modes. We have thus obtained the exact free
actions for the physical modes

\vskip 0.5cm

\fbox{
\begin{minipage}{13cm}
\begin{eqnarray}
S_n^{\, \rm free} &=& C_n \int d^4 x \; Q_n \left[ \, \Box - m_n^2 \,
\right] Q_n \,\,, \nonumber\\ \nonumber\\
C_n &=& \frac{3 \, M^3}{2} \int_0^{y_0} \frac{d y}{A} \, \left[ \frac{3 \,
M^3}{2 \, \phi'^{\,2}} \, \left( A^2 \, {\tilde \Phi}_n \right)'^{\,2} \, 
 + \left( A^2 \, {\tilde \Phi}_n \right)^2 \right] \equiv \frac{1}{2}
\,\,.
\label{normagen}
\\ \nonumber
\\ \nonumber
\end{eqnarray}
\end{minipage}
}

\vskip 0.5cm

The last condition determines the coupling of the scalar modes to
fields on the observable brane. In the next section we discuss this
issue in detail for a couple of specific examples.

\section{Coupling of the perturbations to brane fields}
\label{coupl}

In the previous section we have determined the canonically normalized
scalar modes $Q_n$ which enter in the exact $4$ dimensional
description of the system. Their coupling to SM fields is encoded in
the induced metric $\gamma$ at the observable brane. From the Standard
Model action
\begin{equation}
S_{\rm SM} = \int d^4 x \sqrt{- \gamma} \, {\cal L}_{\rm SM}
\end{equation}
we get
\begin{equation}
S_{\rm int} = - \frac{1}{2} \, \sum_n {\tilde \Phi}_n \left( y_0
\right) \, \int d^4 x \sqrt{- \gamma_0} \, Q_n \left( x \right)
T^\mu_\mu \left( x \right) + \dots \,\,,
\label{int}
\end{equation}
where $T^\mu_\mu$ is the trace of the stress energy tensor of SM
fields, while $\gamma_0$ denotes the background induced metric. Dots
denote higher order couplings, as well as other possible interactions,
as for example the one arising from the coupling of the Higgs boson to
the curvature~\cite{Giudice:2000av}, which will not be considered
here. Hence, the coupling $\Lambda_n$ of the $n-th$ mode is given by
\begin{equation}
{\cal L}_{{\rm int}, n} = \frac{Q_n}{\Lambda_n} \, T^\mu_\mu \quad
\quad , \quad \quad \Big\vert \frac{1}{\Lambda_n} \Big\vert =
\Big\vert \frac{{\tilde \Phi}_n \left( y_0 \right)}{2} \Big\vert \,\,.
\end{equation}
This shows explicitly that the normalization coefficients $C_n$
obtained in the previous section are needed to determine the coupling
of the modes to SM fields.

In the following we compute the mass spectrum and the coupling of the
scalar perturbations for two specific models. As remarked in the
previous Section, we can (with a very good accuracy) neglect the
impact of the brane fields on the wave function of the radion (they
can be treated as probe fields). The first model we consider is a
variation of the Goldberger-Wise mechanism~\cite{GW}, due to DeWolfe
{\it et al.}  (DFGK)~\cite{Dewolfe}, for which the background
equations can be solved analytically. The study of the perturbations
for this model was performed in details in~\cite{csaki2} in the limit
of small backreaction of the scalar field $\phi\,$. In
subsection~\ref{sec:DeWolfe} we extend this study to large
$\phi\,$. The second model, discussed in subsection~\ref{sec:toy}, is
a toy example which does not solve the hierarchy problem. We discuss
it here because both the equations for the background and for the
perturbations can be solved analytically. The two models show similar
features: the mass scale of the radion is given by the backreaction of
the scalar field, whereas the mass scale of the remaining KK modes is
set by the size of the extra dimension. In addition, the coupling of
the radion with SM fields is in both cases stronger than the one of
the KK modes.

\subsection{The DFGK Model}
\label{sec:DeWolfe}

DeWolfe {\it et al.} provided a procedure to ``construct'' bulk and
brane potentials for the scalar field for which the background
equations can be solved analytically~\cite{Dewolfe}. One specific
example studied in~\cite{Dewolfe} is characterized by the following
potentials~\footnote{In the notation of~\cite{Dewolfe}, this
corresponds to the ``superpotential'' $W \left( \phi \right) = 6 \, k
\, M^3 - u \, \phi^2\,$.}
\begin{eqnarray}
\label{eq:pot1}
V &=& -6 k^2 M^3 + \left( 2 \, k \, u + \frac{u^2}{2} \right) \phi^2 -
\frac{u^2}{6 \, M^3} \phi^4 \,\,, \nonumber\\
U_\pm &=& \pm \left( 6 \, k \, M^3 - u \, \phi_\pm^2 \right) \mp 2 u
\phi_\pm \left( \phi - \phi_\pm \right) + \tfrac{1}{2} \mu_\pm \left(
\phi-\phi_\pm \right)^2 \,\,.
\end{eqnarray}
The upper and lower sign refer to the brane at $y=0$ and $y=y_0\,$,
respectively. The quantity $\phi_+$ ($\phi_-$) denotes the value of the
scalar field at the brane at $y=0$ ($y=y_0$). The ``mass parameters''
$\mu_\pm$ do not affect the background solutions (since they do not
contribute to the background junction conditions~(\ref{jump})). However,
they enter in the junction conditions~(\ref{junpert}) for the
perturbations. As we remarked, the $\mu_\pm \rightarrow \infty$ limit
forces $\delta \phi = 0$ at the two branes. For simplicity, the following
calculation is restricted to this limit.

Exact analytical background solutions are computed in normal
coordinate $w$ for the bulk, defined as
\begin{equation}
w = \int_0^y A \left( y \right) d \, y
\end{equation}
(the two branes are located at $w=0\,$ and at $w_0 = w \left( y_0
\right) \,$). The background solutions are
\begin{equation}
\label{eq:sol1}
\phi = \phi_+ {\rm e }^{- u \, w}  \quad \quad , \quad \quad
A = {\rm exp } \left[ - k w - \frac{\phi_+^2}{12 \, M^3} \, \left(
{\rm e }^{- 2 \, u \, w} - 1 \right) \right] \,\,,
\end{equation}
where the normalization $A \left( 0 \right) = 1$ has been
chosen. Following~\cite{csaki2}, we define the parameter $l^2 \equiv
\phi_+^2/ \left( 2 \, M^3 \right)$ to characterize the strength of the
scalar field $\phi$. The $l \ll 1$ limit corresponds to negligible
backreaction of the scalar field on the background geometry. The case
$l = 0$ corresponds to the RS background solution~\cite{RS1}.

The four dimensional Planck mass is given by
\begin{equation}
M_p^2 = 2 M^3 \int_0^{w_0} d w \, A^2 \left( w \right) \simeq 2 M^3
\int_0^{\infty} d w \, A^2 \left( w \right) = \frac{M^3}{k} \, {\rm e
}^{\,l^2/3} \: _1F_1 \left[ \frac{k}{u}, \, \frac{k}{u} + 1, \, -
\frac{l^2}{3} \right] \,\,,
\label{eq:MP}
\end{equation}
where $_1F_1$ is a confluent hypergeometric function. For $u \ll k\,$,
this expression reduces to
\begin{equation}
M_p^2 \simeq \frac{M^3}{k} \quad \quad \quad \left( k \la M \right)
\,\,.
\end{equation}
The electroweak scale at the second brane is generated from the warp
factor
\begin{equation}
A \left( w_0 \right) \simeq \frac{\rm TeV}{M_p} \,\,. 
\label{warp}
\end{equation}

In the present model, the equations for the perturbations can be
solved analytically only in the small $l$ limit. The computation,
summarized in Appendix~\ref{ap2}, gives at leading order in $l$
\begin{eqnarray}
{\rm Radion:} 
&& \Big\vert \frac{1}{\Lambda_{\rm rad}} \Big\vert \simeq \frac{k \,
y_0}{\sqrt{6} \, M_p} \quad,\quad\quad m_{\rm rad} \simeq \frac{2 \, u
\, l}{\left( k \, y_0 \right)^{1+u/k}} \, \sqrt{\frac{2 \, k + u}{3 \,
k}} \,\,\,, \nonumber\\
{\rm KK modes:}
&& \Big\vert \frac{1}{\Lambda_n} \Big\vert \simeq \frac{2 \, \sqrt{2}
\, l \, u}{3 \, n \, \pi \, k} \, \frac{\left(k \, y_0
\right)^{1-u/k}}{M_p} \quad,\quad\quad m_n \simeq \frac{n \, \pi}{y_0}
\quad\quad,\quad\quad n = 1,\, 2,\, \dots
\label{reslowl}
\end{eqnarray}

As we mentioned, the coupling of the radion to the SM is set by the
electroweak scale ($ A \left( y_0 \right) \simeq k \, y_0 \sim
10^{15}- 10^{16}$ sets the hierarchy between the Planck and the
electroweak scales). We also note that the radion mass is proportional
to the backreaction parameter $l\,$, confirming that the radion
corresponds to a flat direction in absence of any stabilization
mechanism (an additional significant suppression to $m_{\rm rad}$ is
given by the $u \left( k \, y_0 \right)^{-u/k}$ prefactor). As a
consequence, the radion mass is expected to be below the physical
cut-off scale for the $y_0$ brane, set by eq.~(\ref{warp})). On the
other hand, the masses for the KK modes are rather insensitive to $l\,$,
and are set by the inverse size ($\sim$ TeV) of the extra
dimension. Finally, we observe that the coupling of the KK modes to
brane fields vanishes in the $l \rightarrow 0$ limit. This was
expected, since in this limit the mode $\Phi$, which is the only one
directly coupled to brane fields in eq.~(\ref{int}), coincides with
the radion.

Eqs.~(\ref{reslowl}) are valid for $l \ll 1\,$. For larger $l$, the
equations for the perturbations cannot be solved
analytically. However, they can be solved numerically, and the
coupling to SM fields can be obtained by numerical evaluation of the
condition~(\ref{normagen}), which is valid for arbitrary background.
We found convenient (i.e., numerics do not develop large hierarchies)
to work in terms of the variable $Y \equiv A^2 \, \Phi\,$, and in
normal coordinate $w\,$. The bulk equation for $Y$ is given in
Appendix~\ref{ap2}. It is supplemented by the junction conditions $Y'
= 0\,$ at the two branes. Hence, we have a boundary value problem,
which we solve by the shooting method: in the numerical integration,
we start from $Y = 1$ (normalization yet to be determined) and $Y' =
0$ at $w =0\,$, choose an arbitrary value for the mass, and evolve the
bulk differential equation for $Y\,$. The integration gives a value of
$Y' \left( w = w_0 \right) \,$ which depends on the choice of
$m\,$. In general, $Y' \left( w_0 \right) \neq 0 \,$, signaling that
we started from a ``bad value'' of the mass $m\,$. Only when $Y'
\left( w_0 \right) = 0$ we have found a physical mode which solves all
the equations for the perturbations.  This occurs only for a discrete
set of $m\,$, which constitute the masses of the physical modes of the
system. A dense scan in $m$ allows to determine the spectrum of the
theory, and the wave function associated to each mode. The
normalization of each wave function is then found by numerical
evaluation of~(\ref{normagen}).

\begin{figure}[h]
\begin{minipage}{16.5cm}
\begin{minipage}[b]{8.25cm}
\begin{flushleft}
  \includegraphics[width=8.25cm]{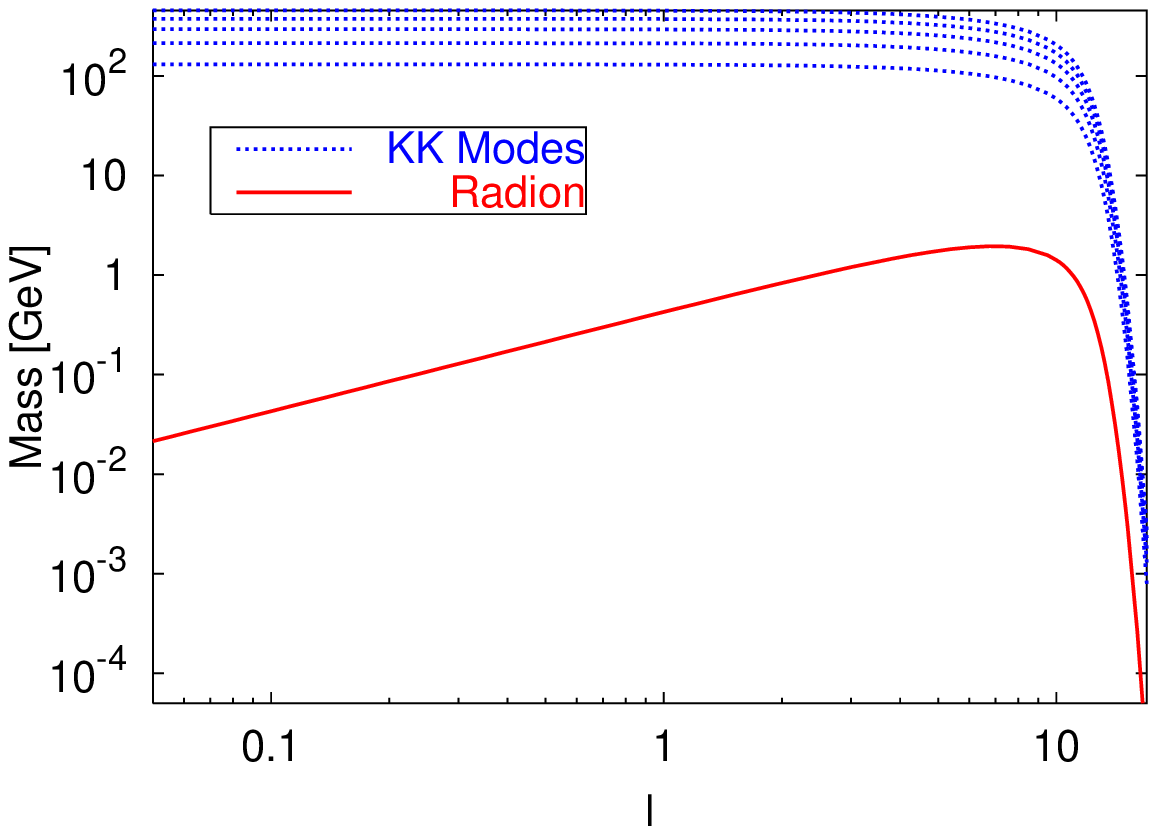}\\
\end{flushleft}
\par
\end{minipage}\hfill
\begin{minipage}[b]{8.25cm}
\begin{flushright}
   \includegraphics[width=8.25cm]{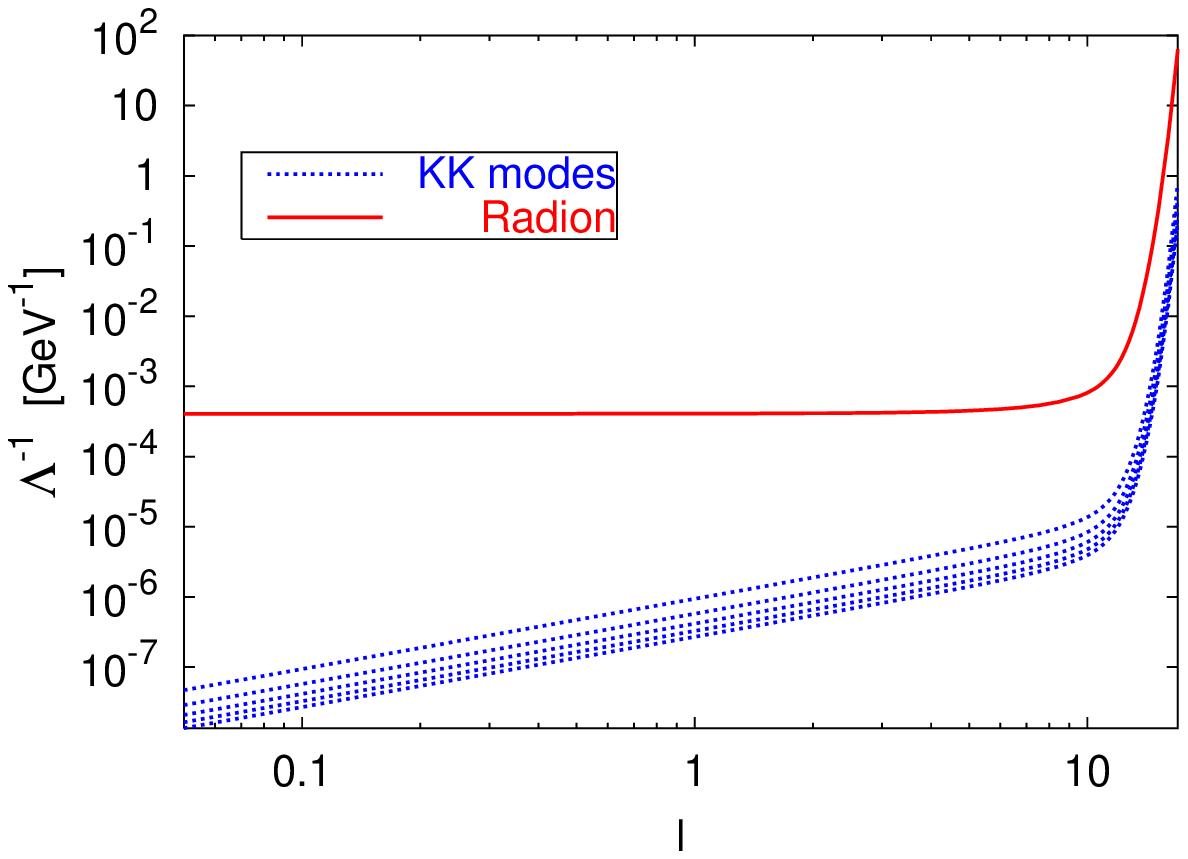}\\
\end{flushright}
\par
\end{minipage}
\end{minipage}
\caption{Mass (left panel) and coupling (right panel) of the radion
and the first KK modes as a function of the (rescaled) amplitude $l$
of the scalar field. Parameters are chosen as $k= 0.1 \, M\,$,
$u=k/40\,$, and in a way to maintain the same warping for all $l\,$.}
\label{fig:k=0.1}
\end{figure}

Figure~\ref{fig:k=0.1} shows the masses (left panel) and the couplings of
the radion and the first (lightest) KK modes as a function of $l\,$.
Besides $l\,$, the system is determined by the four parameter
$M,\,k,\,u\,,$ and $y_0\,$ (or, equivalently, $w_0\,$). We fixed
$u/k=1/40$ (too big or too small values give too small radion masses), and
$k = 0.1 M\,$. The two additional conditions that we need are the recovery
of the Plank mass in the $4$D gravity, eq.~(\ref{eq:MP}), and of the
electroweak scale at the observable brane, eq.~(\ref{warp}). The numerical
results agree with the analytical expressions~(\ref{reslowl}) at low
$l\,$. The radion mass grows linearly with $l\,$, while the modes of the
KK modes are constant. As $l$ increases, the masses show a rather sharp
decrease, which extends also to greater values of $l\,$ than the ones
shown in the Figure. The decrease is probably due to the increase of
$w_0\,$, necessary to preserve a strong warping as $l$ increases (cf.
eqs.~(\ref{eq:sol1}) and~(\ref{warp})). We see that intermediate values
for $l$ are needed, since the radion mass becomes too small both at small
and large $l\,$. The maximum is achieved for $l \la 10\,$, far from the $l
\ll 1$ regime. Also the couplings of the modes to the observable sector,
shown in the right panel, agree with the analytical results at small
$l\,$. The coupling of the radion is constant, while the one of the KK
modes increases linearly with $l\,$ (This behavior has been explained in
the paragraph after eq.~(\ref{reslowl})). The coupling of all the modes
sharply increases at large $l\,$, when the backreaction of $\phi$ cannot
be neglected any longer. Both the decrease of the masses and the increase
of the couplings occurring at large $l$ increase the potential
detectability of the bulk excitations, and stringent experimental bounds
should be expected in this limit.

\begin{figure}[h]
\begin{minipage}{16.5cm}
\begin{minipage}[b]{8.25cm}
\begin{flushleft}
 \includegraphics[width=8.5cm]{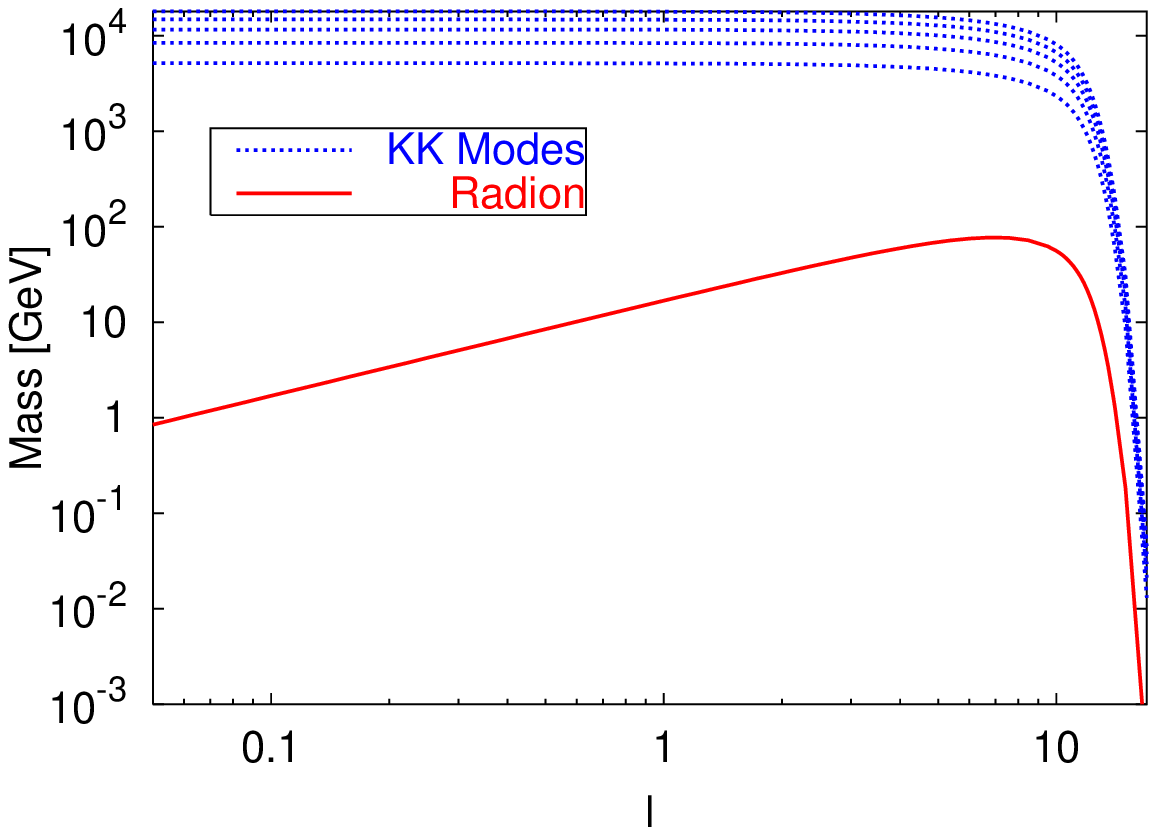}
\end{flushleft}
\par
\end{minipage}\hfill
\begin{minipage}[b]{8.25cm}
\begin{flushright}
  \includegraphics[width=8.5cm]{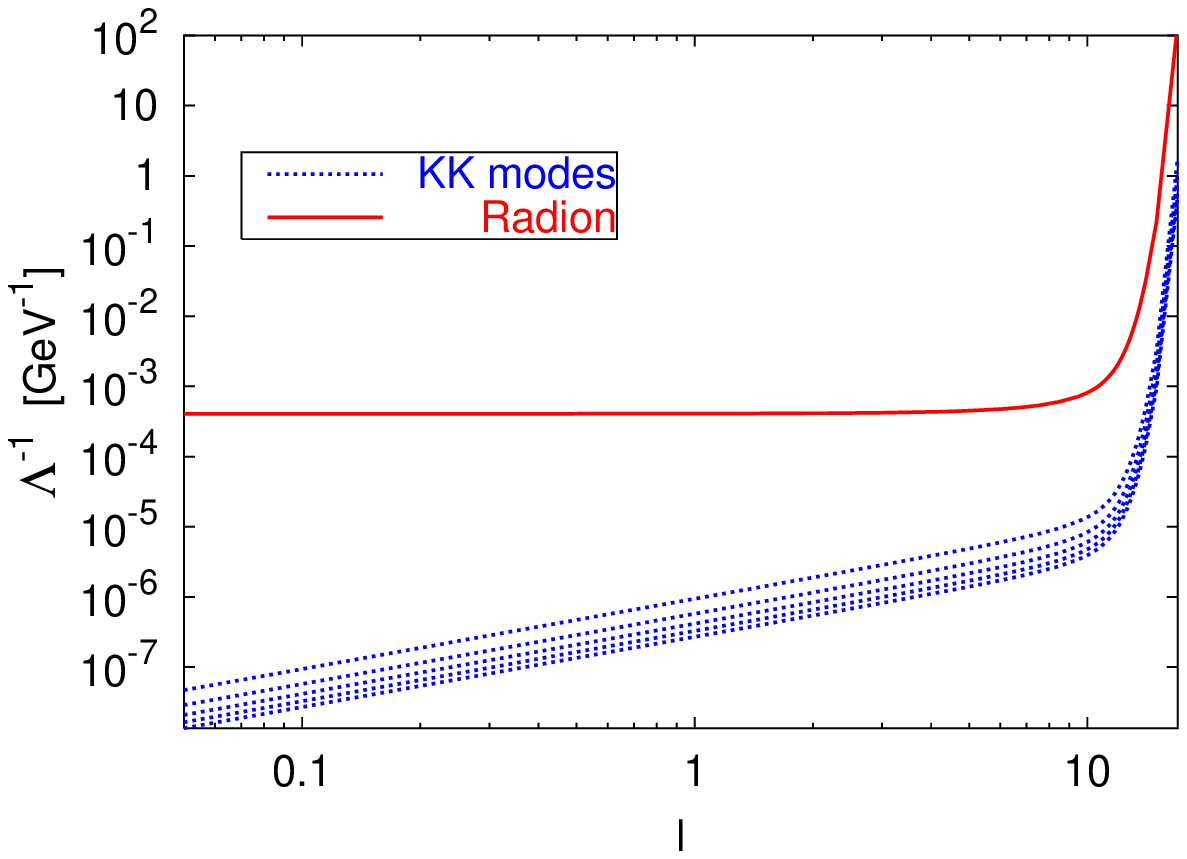}
\end{flushright}
\par
\end{minipage}
\end{minipage}
\caption{Same as in fig.~\ref{fig:k=0.1}, but with $k=M\,$.}
\label{fig:k=1}
\end{figure}

An analogous calculation is summarized in Figure~\ref{fig:k=1}, where
we show the masses and the couplings of the scalar modes for the
choice of parameters $k=M\,$, $u = k/40\,$ ($y_0$ and $M$ are then
fixed as in fig.~\ref{fig:k=0.1}). The numerical results shown in the
two Figures exhibit the same qualitative behavior. The increase of $k$
is responsible for the higher values of the masses shown in
fig.~\ref{fig:k=1}. The values of the couplings are instead very
weakly sensitive on $k\,$, as eqs.~(\ref{reslowl}) indicate.

\subsection{A Toy Braneworld}
\label{sec:toy}

We consider a second example, characterized by the scalar
potentials~\footnote{This is also obtained from the general
prescription given in~\cite{Dewolfe}, using the ``superpotential'' $W
\left( \phi \right) = W_0 \, {\rm exp } \left[ \phi / \left( \sqrt{3
\, M^3} \right) \right] \,$.}
\begin{eqnarray}
V \left( \phi \right) &=& - \, \tfrac{3 \, W_0^2}{24 \, M^3} \, {\rm
exp} \left( \frac{2 \, \phi}{\sqrt{3 \, M^3}} \right) \,\,,
\nonumber\\
U_\pm \left( \phi \right) &=& \mp W_0 \, {\rm exp} \left(
\frac{\phi_\pm}{\sqrt{3 \,M^3}} \right) \mp \frac{W_0}{\sqrt{3 \,
M^3}} \, {\rm exp} \left( \frac{\phi_\pm}{\sqrt{3 \,M^3}} \right) \,
\left( \phi - \phi_\pm \right) + \frac{\mu_\pm}{2} \, \left( \phi -
\phi_\pm \right)^2
\label{toypot}
\end{eqnarray}
(see the previous subsection for the notation). 

The background solutions for this model (given in conformal coordinate
$y$ along the bulk) are
\begin{equation}
\phi = M^{3/2} \, \left[ \alpha \, y + \beta \right] \;\;,\;\; A =
\frac{2 \, \sqrt{3} \, \alpha \, M^3}{W_0} \, {\rm e}^{-\,\frac{\alpha
\, y + \beta}{\sqrt{3}}} \equiv A_0 \, {\rm e}^{- \alpha \, y /
\sqrt{3}} \,\,,
\label{eq:toy_back}
\end{equation}
where $\alpha$ and $\beta$ are arbitrary constants (of mass dimension
$1$ and $0\,$, respectively). Hence, the scalar field $\phi$ is linear
in $y\,$. One can arrange for a strong warping between the two branes,
$\alpha \, y_0 \gg 1\,$. However, it is obtained through a strong
hierarchy (in terms of the parameters entering in~(\ref{toypot}))
between the potentials $U_\pm$ at the two branes. In this sense, the
model does not constitute a natural solution to the hierarchy
problem. However, it is rather instructive, since the bulk equations
for the perturbations of this model can be solved analytically (we
have $z''/z = 3 \, \alpha^2 / 4$ in eq.~(\ref{eqv})).

The junction conditions for the perturbations can be also solved
analytically in the $\mu_\pm \rightarrow \infty$
limit~\footnote{Solutions for finite large $\mu_\pm\,$, can be given
as an expansion series in $1/\mu_\pm$. However, they are not
illuminating, and we will not present them here.}. The mass spectrum
is
\begin{equation}
m_{\rm rad} = \sqrt{\frac{2}{3}} \, \alpha \quad \quad , \quad \quad
m_{\rm KK} = \sqrt{\frac{3 \, \alpha^2}{4} + \left( \frac{n \,
\pi}{y_0} \right)^2} \,\,.
\label{eq:masses2}
\end{equation}
The mass of the radion vanishes in absence of warping, while the KK
modes have a contribution proportional to the warping in addition to
the standard KK mass $\sim n \, \pi / y_0\; \left( n = 1, \, 2, \dots
\right) \,$ . The masses of the radion and of the KK modes become
comparable at large warping, $\alpha \, y_0 \gg 1$. Finally, the
coupling of the different modes to brane fields at the observable
brane are given by
\begin{equation}
\Big\vert \frac{1}{\Lambda_{\rm radion}} \Big\vert =
\frac{\alpha^{1/2}}{2^{1/2} \, 3^{3/4} \, M^{3/2} \, A_0^{3/2}} \,
{\rm e}^{\sqrt{3} \, \alpha \, y_0 / 2} \; \; , \; \; \Big\vert
\frac{1}{\Lambda_{\rm KK}} \Big\vert = \frac{4 \, n \, \pi \,
\alpha^{1/2}}{3 \, M^{3/2} \, A_0^{3/2}} \frac{1}{\left( \alpha \, y_0
\right)^{3/2}} \, {\rm e}^{\sqrt{3} \, \alpha \, y_0/2} \,\,.
\label{eq:coup2}
\end{equation}
As in the previous example, the radion is more strongly coupled than
the other KK modes, although the couplings have comparable strength.

\section{Numerics of the perturbations with the BraneCode}
\label{branecode}

The analytical results derived above can be be supported by very
different means, namely by the numerical study of the braneworld
dynamics. Suppose we have a stable flat brane configuration plus
fluctuations around it. This can be achieved by taking initial
conditions which are slightly displaced from the stable configuration,
so that the system will start oscillating around it. In this way we
can study the properties and the time evolution of the perturbations,
and see what relation they have with the eigemodes of the system
computed analytically, eq.~(\ref{normagen}). The evolution of the
interbrane distance $D \left( t \right)$ is often characterized by
several oscillatory modes, superimposed to a lower frequency
oscillation. Correspondingly, the Fourier transform of $D \left( t
\right)\,$ shows the presence of resonance bands: the one with lowest
frequency is identified with the radion field, while the other ones
with the higher KK modes. The numerical values of the frequencies of
the bands can be compared with the analytical expression for the
masses of the modes. In addition, we can split the time evolution in
successive time intervals (each containing several oscillations), and
perform the Fourier transform in each of them. This allows to study
how the excited modes evolve in time.

We make use of a numerical code, named {\it BraneCode}~\cite{bc},
designed for the study of the broad class of brane models~(\ref{ac1})
we are interested in. The numerical integration is performed with the
assumption of homogeneity and isotropy along the brane coordinates, so
that the system solved is effectively two-dimensional (time and bulk
coordinates). This limitation removes the tensor modes from the
numerical evolution (since there are no tensor modes in two
dimensions), which is welcome for the present analysis. However, it
also restricts the study of the scalar modes to the large wavelengths
limit.  Since we numerically integrate the exact $5$D Einstein
equations, the present study goes beyond the linear regime for the
perturbations, which limits instead the analytical calculations. In
the example shown below, the nonlinear dynamics is relevant at the
earlier stages of the evolution.

Some attention has to be paid to the choice of the initial conditions,
since they have to respect two constraint (nondynamical) Einstein
equations. The simplest possibility is to search for static bulk
configurations, with
\begin{equation}
d s^2 = A \left( y \right)^2 \left[ - d t^2 + {\rm e}^{2 H t} \, d
{\bf x}^2 + d y^2 \right] \,\,,
\end{equation}
and with $\phi = \phi \left( y \right) \,$. As discussed in~\cite{bc},
the class of potentials considered in section~\ref{coupl} admits at
most two static configurations. When they exist, they are
characterized by different values of $H\,$, which is proportional to
the Hubble parameter measured by brane observers. The solution with
higher $H$ is unstable, while the other one is stable. As explained
in~\cite{FK}, a non-vanishing $H$ gives a negative contribution to the
physical masses of the scalar fluctuations around the solution,
$m_{\rm tot}^2 \leq m_0^2 - 4 \, H^2 \,$. In many cases $m^2_{\rm tot}$
is negative, meaning that the corresponding static configuration
is subject to a tachyonic instability~\cite{FK}. The instability
typically leads towards the stable configuration, characterized by the
lower expansion rate~\cite{bc}.

\begin{figure}[h!]
\centering
\includegraphics[width=16.5cm]{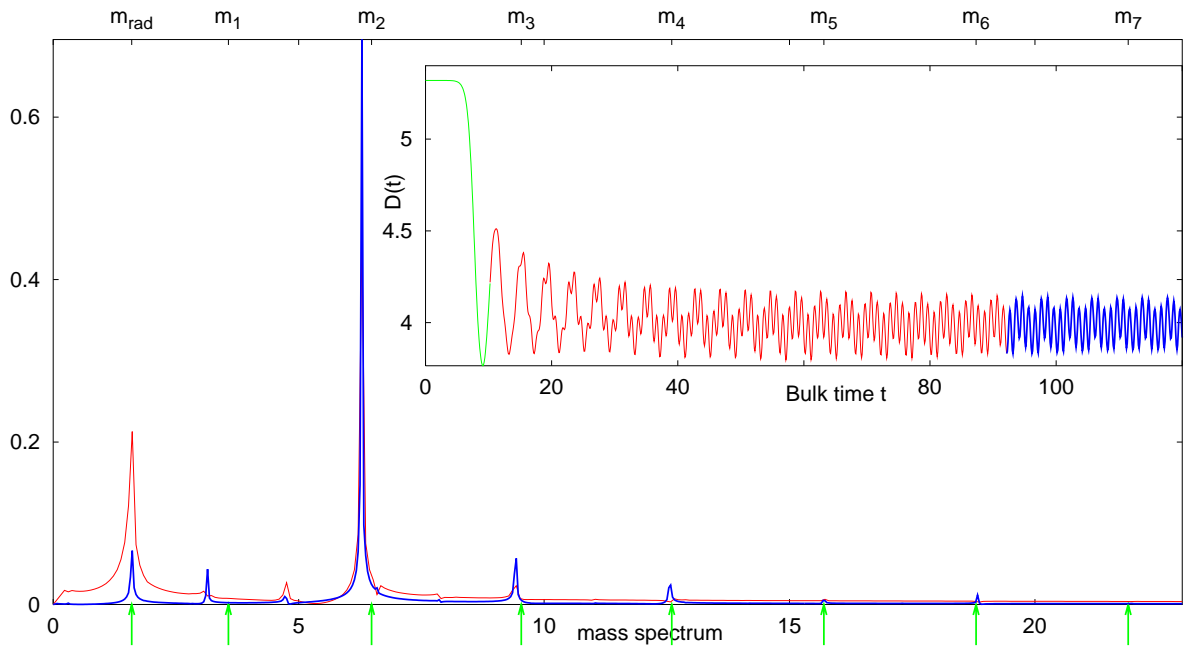}
\caption{Relaxation towards a stable braneworld configuration in the
model~(\ref{toypot}). The evolution of the interbrane distance $D \left( t
\right)$ and its Fourier transform are shown. The arrows beneath the
frequency axis denote the masses of the radion and of the first KK modes
according to eq.~(\ref{eq:masses2}). Parameters are chosen as $\alpha \,
y_0 = 2 ,\, \beta = 0 , \, {\rm e}^{-\alpha \, y_0/\sqrt{3}} \, A_0 \,
\mu_+ = A_0 \, \mu_- = 100 \, M \,$, and the figure is shown in units
$M=1\,$.}
\label{fig:relax2}
\end{figure}

A systematic algorithm for the search of static configurations is
described in~\cite{bc}. In the present discussion, we focus on the
model~(\ref{toypot}), and start from an unstable static solution.
Figure~\ref{fig:relax2} shows the relaxation of this configuration
towards a stable solution, of the class discussed in the previous
section. We plot the time evolution of the interbrane distance,
defined as
\begin{equation}
D \left( t \right) = \int_0^{y_0} d y \, A \left( t,y \right) \,\,,
\label{eq:interbrane}
\end{equation}
together with its Fourier transform. We show two different spectra,
according to different (consecutive) time intervals immediately after
the ``transition'' between the two solutions.

The numerical result confirms the rapidity of the transition,
and the initial excitation of the radion mode. The subsequent
evolution can be roughly divided into two stages. During the
earlier one, we can see the excitation of modes of higher frequency
superimposed to the radion oscillation. The positions of the modes are
in excellent agreement with the masses obtained analytically - see
eq.~(\ref{eq:masses2}) - which are indicated by the arrows below the
frequency axis. These confirm that the modes excited are precisely the
ones found through the linearized analytical calculation. We note that
the excitation of the higher modes starts already during the first
oscillation of the radion field. It is also possible to see a decrease
of the amplitude of the radion oscillations during all this first
stage. The second stage is instead characterized by a constant pattern
of oscillatory modes. The spectrum of the perturbations does not
change appreciably during these later times.

These features can be readily understood. The large amplitude of the
radion oscillations is due to the strong tachyonic instability of the
first configuration. The excitation of the higher frequency modes is
due to the coupling of the perturbations which occurs at the nonlinear
level (as we mentioned, the numerical integration automatically takes 
into account the nonlinear dynamics of the system). Most of the energy
of the radion field is transferred to the higher modes, as also
indicated by the decrease of the overall amplitude occurring in the
first stage. At later times, the amplitudes of the oscillations become
sufficiently small so that nonlinear effect can be neglected. The fact
that the patterns of the oscillations become constant confirms that
the modes are decoupled at the linear level, as eq.~(\ref{normagen})
indicates.

In the present computation, we did not include any brane fields. Their
coupling to the radion may be responsible for some energy transfer
from the bulk to fields on the brane, during the earlier stages of
cosmological evolution. This possibility is discussed in greater detail
in the discussion section.

\section{Discussion} \label{discussion}

We considered $5$D braneworld models with the inclusion of a bulk
scalar field for the sake of generality and, more importantly, to
provide stabilization of the extra dimension. Observers on a $4$D
brane embedded in this space will in general experience corrections to
the $4$D Einstein gravity, as well as additional four dimensional
degrees of freedom emerging from the bulk gravity and the bulk scalar.
These excitations interact with the fields of the Standard Model (or
more generically, of the observable sector) living on our brane.
Therefore, their nature and the details of their couplings to the
observable sector are the subject of particle physics phenomenology
and experiments. The set-up we studied is a specific example of more
general systems characterized by a higher dimensional warped bulk
geometry, which emerges not only in the context of $5$D braneworlds,
but also in many other models motivated by string cosmology, see
e.g.~\cite{KKLT}. For this reason we will comment not only on the
results of the explicit calculations of the paper, but also on
potential extensions of our methods and results for the broad context
of the warped geometries.

In the present paper we focused on the scalar excitations of the
system. The warp bulk geometry can be strongly curved, and the full
machinery of General Relativity is required for the study of the bulk
gravity/scalar dynamics. In general, the physical modes associated
with the excitations of the coupled gravity/scalar system do neither
coincide with the fluctuations $\delta \phi$ of the scalar field nor
with the scalar perturbations $\Phi$ of the metric, but rather with
some linear combination of them. This is sharply different from the
identification of the eigenmodes in the familiar KK dimensional
reduction, where higher dimensional metric components decouple from
the bulk fields. A much more similar context is the one of $4$D
cosmology with scalar fields, where cosmological scalar perturbations
of the metric are sourced by the fluctuations of the inflaton. In the
present paper we extended this approach to the more complicated set-up
of higher dimensional warped geometries.

This analysis can be pursued in several directions.

1. We computed the coupling of the radion (the lightest eigenmode of
the warped geometry excitations) and of the higher KK modes to
Standard Model particles, which is one of the most crucial pieces of
information for phenomenological studies. The physical modes emerge
from the decomposition and diagonalization of the second order action
for the perturbations. The diagonalization requires great care, since
the starting action for the perturbations is particularly involved,
and the identification of the relevant dynamical variables is far from
being obvious. Contributions both from the bulk and the branes are
present in this action, neither of which can be immediately
diagonalized. In particular, {\it a priori} one could have wondered
whether the two different contributions could have been simultaneously
diagonalized, or if some mixing between the different KK levels would
have unavoidably remained.

We performed this tedious decomposition in the case of $5$D braneworlds
with bulk scalar field and two orbifold branes at the edges. The answer
happens to be rather transparent and simple: the total action for the
perturbations from the bulk and the two branes can be diagonalized in
terms of free, non-interacting scalar KK excitations. The canonical
quantization of the actions of these modes specifies the normalization of
their wave function along the extra dimension. This in turns defines the
value of the perturbations at the observable brane, which sets the coupling
between the scalar modes and the SM fields (the overall normalization
cannot be obtained from the $5$D Einstein equations for the perturbations,
which are all linear in the perturbations). For the reasons just remarked,
it should not be a surprise that the couplings are very sensitive on the
details of background bulk geometry and of the bulk scalar field
distribution. We illustrated this dependence with some specific examples,
see for example Figs.~\ref{fig:k=0.1} and~\ref{fig:k=1}.

One of the examples we have considered is related to the
Goldberger-Wise mechanism~\cite{GW} for the stabilization of the
Randall-Sundrum model~\cite{RS1}. More specifically, we consider the
model by De Wolfe et al. (DFGK)~\cite{Dewolfe}, where the background
equation for the stabilized geometry and the bulk scalar field can be
solved analytically.  It is convenient to introduce a parameter
$l$~\cite{csaki2} - proportional to the amplitude of the scalar field
- which characterizes the departure of the model from the pure
Randall-Sundrum model without stabilization. Thus, $l=0$ corresponds
to the Randall-Sundrum model, small $l \ll 1$ to small backreaction of
the background scalar on the bulk AdS geometry, while large $l \ga 1$
implies a significant modification of the geometry due to the presence
of $\phi\,$. It is important to bear in mind that a large warping
(large difference of the background conformal factor $A$ between the
two branes) can be obtained for all $l\,$. Thus, the virtue of the
braneworld models to solve the hierarchy problem is preserved also for
cases, characterized by large $l\,$, in which the bulk geometry may
significantly differ from the AdS geometry of~\cite{RS1}. Indeed, in
the specific example we have considered we were able to preserve the
same warping for all values of $l\,$.

We shall make comment on the physical reliability of the models with
small vs. non-small $l$. It is not easy to accept that the bulk scalar
field gives only negligible, perturbative contribution to the
background bulk geometry, and that its only role is to provide a tiny
lift of the radion mode, without affecting the geometry in any other
way. As we already remarked, the idea of RS is to avoid strong
hierarchies besides the one given by the warp factor. If $l$ is
limited to the small values needed to remain in the perturbative
regime, the radion mass turns out to be significantly smaller than the
electroweak scale. Although this is certainly a much milder hierarchy
than the one between the electroweak and the Plank scales, it can be
avoided provided $l$ is sufficiently large. Another point in favor of
a large $l$ is related to the dynamics of the radion in the early
universe. To appreciate this, we computed and showed in
Appendix~\ref{ap3} the $4$D effective potential for the radion (this
effective potential is only approximately valid at small $l\,$, and
should not be confused with the exact $4$D description given in the
main text). In terms of the canonically normalized field
(corresponding to the radion in this effective description), the local
minimum at which the stabilization occurs is minuscule. If we are
interested in the early dynamics of the braneworld, it is very hard to
understand how the system can go to such a minimum, unless it starts
extremely close to it. This is a very similar problem to the one of
moduli stabilization in heterotic string cosmology~\cite{bs}. This
problem is enhanced for small $l\,$. These reasons are a strong
motivation for studying the system also beyond the small $l$ limit.

The coupling of the radion to SM fields was already computed by Csaki
et al. \cite{csaki2} in the limit of small backreaction, $l \ll
1\,$. Our formula reproduces their result in this limit. In addition,
we found that the strength of the interaction remains almost constant
in the range $0< l \la 10$.  Most interestingly, however, is that for
$l > 10$ the value of the coupling begins to grow sharply as a
function of $l$.  Therefore we conclude that, in the model considered,
a stronger stabilization can enhance even by orders of magnitude the
coupling of the radion to SM fields.

Another very interesting property we found is the strong dependence of the
masses of the radion and of the KK modes on the parameter $l\,$. In the
range $0 < l \la 10$ the mass of the radion increases linearly with $l$
(consistently with the fact that the radion becomes massless as $l
\rightarrow 0\,$), while the masses of the KK modes are nearly constant.
At $l \ga 10$ the masses of all the modes are instead sharply decreasing.
The decrease is probably due to the larger interbrane separation which
must be imposed at large $l\,$ - in order to preserve a strong warping at
the observable brane - and which results in lower masses from gradients in
the $y$ direction. Hence, both very small and large $l$ are
phenomenologically excluded, since they lead to a too light radion. The
radion mass is maximized for $l \sim 10\,$, far from the perturbative $l
\ll 1$ regime. This paper provides the technical tools for extending the
phenomenological studies to this relevant regime.

2. We identified the KK eigenmodes of the braneworld excitations in the
general case, without any assumption of small backreaction of the bulk
scalar field. In the linear regime they are free non-interacting
dynamical fields, since they correspond to canonically normalized
eigenmodes of the action expanded up to second order in the
perturbations. However, we may expect that they are coupled at higher
(third and further) order in the decomposition of the full action. To
verify this, we made use of numerical simulations of the dynamical
evolution of the braneworlds, which give complementary informations to
the analytical study. The simulations are based on the BraneCode,
which was developed to solve the fully nonlinear self-consistent
evolution of the braneworld with the scalar field \cite{bc}. We
started with an unstable brane configuration, which evolves to a
stable one plus excitations around it. The excitations are composed
of a superposition of oscillatory modes, whose frequencies (determined
by a Fourier decomposition) coincide with the eigenmasses found by the
analytical study of the linear regime. The higher KK modes are excited
by the nonlinear interactions with the radion field during the earlier
stages of the evolution. At later times, the nonlinear interactions
can be neglected, and the pattern of oscillations becomes constant.

The discussion of the brane system restructuring and of the excitation of
its eigenmodes leads us to another subject where the interaction of the
radion and the other KK modes with SM particles is important. This is
(p)reheating of the universe after inflation, and, in particular, the role
that the brane excitations may have had in the generation of brane fields.
Inflation in the braneworld setting corresponds to the curved brane, so
that after inflation the brane is flattening as the brane curvature
decreases with the decrease of the Hubble parameter $H(t)$. As we just
noted, the reconfiguration of the brane geometry leads to interbrane
oscillations, which correspond to excitations of the radion and the other
KK modes. The interaction of the bulk modes with the SM particles will
results in the decay of these oscillations. More specifically, let us
consider a coherently oscillating radion $\Phi=\Phi_0 \, \cos {m_{rad} \,
t}$ interacting with some scalar field $\chi$ of mass $m$ living on our
brane. The equation for the amplitudes of the eigenmodes $\chi_k(t) e^{i
\vec k \vec x}$ of quantum fluctuations of $\chi$ can be reduced to
\begin{equation}
\label{modes} \ddot \chi_k +\left( k^2 +m^2-(m^2+ \frac{1}{2} m^2 _{rad})
\Phi_0 \, \cos {m_{rad} \, t} \right) \chi_k = 0
\end{equation}
(more accurately, $\chi_k$ refers to the canonically normalized field,
and $m$ to the physical mass, both obtained after a rescaling with the
conformal factor at the brane; in addition, one should take into
account that the amplitude $\Phi_0$ of the oscillations is decreasing
due to the cosmological expansion of the brane). This is the familiar
equation for the parametric resonant amplification of quantum
fluctuations interacting with the oscillating background field
\cite{KLS94}. The strength of the effect is given by the parameter
$q=(\frac{m^2}{m^2 _{rad}} + \frac{1}{2}) \, \Phi_0$. The amplitude of
the radion oscillations (here $\Phi_0$) corresponds to the induced
metric fluctuation, which is below unity. Thus, in case of a mass
ratio $\frac{m^2}{m^2 _{rad}} < 1\,$, the parametric resonance is not
strong enough for preheating. In the opposite case, $\frac{m^2}{m^2
_{rad}} \gg 1 \,$, the parameter $q$ is large, but the effect works
only in the higher instability bands where $k^2 > \frac{m^2}{m^2
_{rad}}\,$, and it is again very small. We conclude that it is hard to
have preheating effects from the radion oscillations, so that the
radion oscillations decay in the perturbative regime. An interesting
earlier paper~\cite{coll} seems to conclude that preheating effects
associated to the oscillations of the radion are instead more
significant. We think that, once the expansion of the brane, and the
consequent decrease of the amplitude of the oscillations, are taken
into account, the result claimed in~\cite{coll} may be significantly
reduced.

3. Finally, we will discuss other virtues of the method of the full
action series decomposition which we used in the paper. While we focus
on the problem of identification of the eigenmodes of the
gravity/scalar dynamical system in order to quantize the radion and
the higher KK modes, in principle we can extend this method to discuss
another important sector of the effective $4$D theory, namely
effective $4$D gravity. There is extensive literature on the
properties of the $4$D gravity in the braneworld models; here we
discuss only the low energy limit, that is the linearized gravity at
the brane \cite{cgr,shift,Mukohyama:2001ks}. On general grounds, one
expects that the $4$D effective gravity at the brane will be a
Brans-Dicke (BD) theory. For pure RS model without brane
stabilizations, the BD gravity contains a massless scalar, which is
the radion. For stabilized branes the radion acquires a mass, so that
one expects a BD theory with a massive BD scalar. The latter
situation was studied in~\cite{R}, where the Lagrangian for the $4$D
gravity was derived as a series containing the Einstein term $R$ plus
higher derivatives term of the form $R^2/m^2_{rad}$, plus even higher
derivatives terms.

The main method adopted in the literature for the derivation of the
low energy $4$D effective gravity is based on the introduction of a
point-like stress-energy source at the visible brane, and on the
investigation of the gravitational response upon this source,
including the bulk gravity, the scalar field, and the shift in the
embedding of the branes~\cite{shift}. Notice, however, that the
signature of the low energy gravity (BD or $R^2$ theory) is manifested
in the free gravitational modes even without the stress-energy
sources. Therefore, in principle we can address the question regarding
brane gravity in terms of the series decomposition of the full action.
To do so, we need to include not only the scalar metric perturbations
$\Phi$ in the form (\ref{line}), but also the transverse-traceless
gravitational modes $h^{TT}_{AB}$. Bulk gravitational waves in general
have $5$ degrees of freedom. However, the massless scalar and vector
projections of the bulk graviton are absent, so that the massless
graviton has only $2$ degrees which corresponds to usual $3+1$
dimensional transversal and traceless (TT) gravitational waves
$h^{TT}_{\mu\nu}\,$~\cite{andrei}. We can extend (\ref{line}) to the
form where the full $5$D metric perturbations $h_{AM}$ are
\begin{equation}\label{gauge}
h_{55}=2\Phi \ , \,\,\, h_{5\mu}=0 \ , \,\,\,
h_{\mu\nu}=h^{TT}_{\mu\nu}-\eta_{\mu\nu} \Phi \ .
\end{equation}

We then have to include TT modes in the decomposition of section
\ref{sec:2A}. Up to the second order in perturbation series, the
effective $4$D action acquires the new term $h^{TT}_{\mu\nu;\rho}
h^{TT;\rho}_{\mu\nu} / 4$ which is equal to the linearized $4$D
curvature $R$, see e.g. \cite{andrei}. The BD coupling $(1+Q) R$
emerges only in the third order of the perturbation series, which is
beyond the scope of this paper. Here we just note that a terms with
the structure $\Phi h^{TT;\rho}_{\mu\nu} h^{TT}_{\mu\nu;\rho}$ will
appear in the third order decomposition.

What is important, is that with the present approach we explicitly see
that the BD scalar is a massive one. This follows from the second
order decomposition of section~\ref{sub2}. Now, how the BD theory with
the massive radion (as a BD scalar) can be reconciled with the $R+R^2$
theory derived in \cite{R}?  It is well known \cite{conf} that a
theory of gravity with the higher derivatives in the form ${\cal
L}_1=-\frac{M_p^2}{16\pi} \left( R+\frac{R^2}{6m^2}\right)$ is
conformally equivalent to the Einstein theory with the scalar field
$\phi$ and with Lagrangian ${\cal L}_2=-\frac{M_p^2}{16\pi} R
+\frac{1}{2} \phi_{\mu} \phi^{\mu} -V(\phi)$. The metrics of theories
${\cal L}_1$ and ${\cal L}_2$ are related by conformal transformation
$g_{\mu\nu} \to \left(1- \frac{R}{3m^2} \right) \, g_{\mu\nu}$.  The
scalar fields $\phi$ is obtained from the scalar curvature $R$ as
\begin{equation}\label{scalar}
\phi=\sqrt{\frac{3M_p^2}{16\pi}} \ln \left(1- \frac{ R}{3m^2} \right)
\,\,,
\end{equation}
and the potential is given by formula
\begin{equation}\label{pot}
V(\phi)= \frac{3M_p^2 m^2}{32\pi} \left[1-
\exp\left(-\sqrt{\frac{16\phi}{3M_p^2}} \phi \right) \right] \ .
\end{equation}
In the limit of small $\phi$ we obtain $V(\phi) \approx \frac{1}{2}m^2
\phi^2$.  Thus, for small values of the radion (i.e. BD scalar) the
Einstein gravity with small high derivative corrections is equivalent
to the theory without $R^2$ term but with instead the additional
massive scalar.

\section*{Acknowledgments}

We are grateful to Robert Brandenberger, Carlo Contaldi, Gia Dvali, Antony
Lewis, Slava Mukhanov, Shinji Mukohyama, Dimitri Podolsky, Erich Poppitz,
Misao Sasaki, and Lorenzo Sorbo for fruitful discussions. In particular,
we thank Csaba Csaki for important comments and clarifications on the
work~\cite{csaki2}.

\appendix

\section{Computation of $S_2$} \label{ap1}

In this Appendix we outline the computation of the action~(\ref{ac2})
for the perturbations. Following~\cite{mfb}, we first perform an ADM
decomposition along $y$ of the gravitational part of the
action~(\ref{ac1}).  One can easily verify that, for a line element of
the form
\begin{equation}
d s^2 = {\cal A}^2 \left( t, {\bf x}, y \right) \left( - d t^2 + d {\bf x}^2
\right) + {\cal B}^2 \left( t, {\bf x}, y \right) d y^2 \,\,,
\end{equation}
one has
\begin{equation}
\sqrt{\vert G \vert} R = \sqrt{\vert G \vert} \left\{ R_{ \left( 4 \right)}
+ \frac{12 \, {\cal A}'^2}{{\cal B}^2 \, {\cal A}^2} \right\} - 8 \left[ 
\frac{{\cal A}^3 \, {\cal A}'}{{\cal B}} \right]' - 2 \, \eta^{\mu \nu}
\partial_\mu \left( {\cal A}^2 \, \partial_\nu {\cal B} \right) \,\,,
\label{adm}
\end{equation}
where $R_{ \left( 4 \right)} = - 6 \eta^{\mu \nu} \left( \partial_\mu
\partial_\nu {\cal A} \right) / {\cal A}^3$ is the curvature of the
four dimensional slices. The last term in~(\ref{adm}) is a boundary
term on the $3+1$ slices, and it can be dropped. On the other hand, the
total derivative in $y\,$ does not vanish, since the space in that
direction is limited by the two branes. However, its contribution to
the action precisely cancels with the Gibbons-Hawking term
$\sqrt{-\gamma} \left[ K \right]$ in the starting action~(\ref{ac1})
for the two branes.

Hence, only the first term of~(\ref{adm}) is relevant. Combining it with
the bulk action for the scalar field, we get
\begin{equation}
S_{\rm bulk} = 2 \int_0^{y_0} d^5 x {\cal A}^4 \, {\cal B} \left[ - 6
\, M^3 \, \eta^{\mu \nu} \frac{\partial_\mu \partial_\nu {\cal
A}}{{\cal A}^3} + \frac{12 \, M^3 \, {\cal A}'^2}{{\cal B}^2 \, {\cal
A}^2} - \frac{\eta^{\mu \nu}}{2 \, {\cal A}^2} \partial_\mu \varphi \,
\partial_\nu \varphi - \frac{1}{2 \, {\cal B}^2} \varphi'^2 - V
\right] \,\,.
\end{equation}

The coefficients ${\cal A}$ and ${\cal B}$ are obtained from the line
element~(\ref{line}). Expanding the bulk action at second order in the
perturbations, and making use of the background equations to simplify
the final expression, we obtain the bulk part of the action~(\ref{ac2}).
The brane contributions to~(\ref{ac2}) are instead the
expansion of the last term of~(\ref{ac1}).

\section{Perturbations in the DFGK model for $l \ll 1$}
\label{ap2}

In the following, we derive the analytical results given
in~(\ref{reslowl}), valid in the $l \ll 1$ regime. Some of the present
results were also obtained in~\cite{csaki2}. It is useful to work in terms
of the rescaled mode ${\tilde Y}_n = A^2 \, {\tilde \Phi}_n\,$, and in
normal coordinate $w\,$. By combining the equations for the background
quantities and the perturbations, one finds
\begin{equation}
{\tilde Y}_n'' - 2 \left( \frac{\phi''}{\phi'} + \frac{A'}{A} \right)
{\tilde Y}_n' + \left[ 2 \, \frac{A''}{A} - 2 \, \frac{A'^2}{A^2} +
\frac{m_n^2}{A^2} \right] {\tilde Y}_n = 0 \quad \quad , \quad \quad
{\tilde Y}_n' |_{w=0,w_0}=0 \label{eqy}
\end{equation}
(in this Appendix, prime denotes differentiation with respect to normal
coordinate $w\,$).

For the moment, we restrict the computation to $l=0\,$, that is we
ignore the backreaction of the scalar field on the background geometry
(notice that $\phi''/\phi' = - u$ is regular in this limit). The bulk
equation is then solved by
\begin{equation}
{\tilde Y}_n = {\tilde Y}_{n}^{(0)} \, {\rm e}^{-w \left( k + u
\right)} \left[ J_{1+u/k} \left( \frac{m_n}{k} \, {\rm e}^{k w}
\right) + {\cal C}_n \, Y_{1+u/k} \left( \frac{m_n}{k} \, {\rm e}^{k
w} \right) \right] \,\,,
\label{modesol}
\end{equation}
with arbitrary constants $ {\tilde Y}_{n}^{(0)}$ and ${\cal
C}_n\,$. Differentiating this solution gives
\begin{equation}
{\tilde Y}_n' = - m_n \, {\tilde Y}_{n}^{(0)} \, {\rm e}^{-w \, u } \left[
J_{2+u/k} \left( \frac{m_n}{k} \, {\rm e}^{k w} \right) + {\cal C}_n \,
Y_{2+u/k} \left( \frac{m_n}{k} \, {\rm e}^{k w} \right) \right] \,\,.
\label{diffsol}
\end{equation}
The boundary condition at $w = 0$ can be used to enforce
\begin{equation}
{\cal C}_n = - \, \frac{J_{2 + u/k} \left( \frac{m_n}{k} \right)}{Y_{2
+ u/k} \left( \frac{m_n}{k} \right)} \rightarrow \frac{\pi}{\Gamma
\left( 2 + \frac{u}{k} \right) \, \Gamma \left( 3 + \frac{u}{k}
\right)} \, \left( \frac{m_n}{2 \, k} \right)^{4 + \frac{2\,u}{k}}
\,\,,
\label{yc}
\end{equation}
where the last limit holds for $m_n/k \rightarrow 0\,$. As we will now
see, this is the case we are interested in, since $m$ is set to the
electroweak scale, while $k \sim M_p \,$.

Eq.~(\ref{diffsol}) shows that a mode~(\ref{modesol}) with vanishing mass
solves also the boundary conditions. This mode corresponds to the radion,
whose mass is due to the backreaction of $\phi$ on the geometry: a
non-vanishing radion mass is found only once the bulk equation~(\ref{eqy})
is expanded at the first nontrivial order in $l\,$. Let us postpone this
calculation till the end of this Appendix, and concentrate on the KK
modes, for which the present calculation ($l=0$) is accurate enough. As
clear from the expansion~(\ref{yc}), the second term in~(\ref{modesol}) is
completely irrelevant for $m_n / k \sim {\rm TeV} / M_p \,$. The junction
condition at $w = w_0$ then gives
\begin{equation}
J_{2 + \frac{u}{k}} \left( \frac{m_n}{k} \, {\rm e^{k \, w_0}} \right)
\simeq 0 \,\,.
\end{equation}
Hence, the masses of the KK modes are determined by the poles of this
function. By expanding it for large argument we find
\begin{equation}
m_n \simeq \frac{\pi}{y_0} \left[ n + \frac{3}{4}+ \frac{u}{2 \, k}
-\frac{15+16u / k + 4 u^2 / k^2}{8 \, \pi^2 \, n} + {\cal O}
\left(n^{-2} \right) \right] \quad,\quad\quad n=1,\,2,\,3,\,4,\,5,\dots
\label{eq:mapprox1}
\end{equation}
This approximation becomes more and more accurate as $n$
increases. However, the comparison with numerical results shows very
good agreement at all $n\,$. The normalization coefficient ${\tilde
Y}_n^{(0)}$ is computed by inserting the solution~(\ref{modesol}) into
the normalization condition~(\ref{normagen}). Since we are interested
in the leading result for small $l\,$, only the first term
in~(\ref{normagen}) has to be computed. We get
\begin{equation}
\frac{9 \, M^3 \, m_n^2 \, {\tilde Y}_n^{(0) \: 2}}{8 \, l^2 \, u^2}
\, \int_0^{w_0} d w \, {\rm e}^{2 \, k \, w} \, J_{2+\frac{u}{k}}^2
\left( \frac{m_n}{k} \, {\rm e}^{k \, w} \right) \simeq \frac{9 \, M^3
\, {\tilde Y}_n^{(0) \: 2}}{16 \, k \, l^2 \, u^2} \, m_n^2 \, {\rm
e}^{2 \, k \, w_0} \, J_{1+\frac{u}{k}}^2 \left( \frac{m_n}{k} \, {\rm
e}^{k \, w_0} \right) \equiv \frac{1}{2} \,\,.
\end{equation}
We can now evaluate the normalized ${\tilde \Phi}_n = {\tilde Y}_n /
A^2$ at the $w_0\,$ brane. We obtain, at leading order, the coupling
of the KK modes with brane fields reported in eq.~(\ref{reslowl}).

We still have to compute the normalization and the mass of the radion
field. As remarked, the mass can be obtained by inserting the
background solutions~(\ref{eq:sol1}) into the bulk eq.~(\ref{eqy}),
and expanding at the first nontrivial order in $l\,$. The
normalization can however be found already from the $l=0\,$ term. At
zero-th order in $l\,$, the radion is massless, and eqs.~(\ref{eqy})
are solved by a constant ${\tilde Y}\,$ (as it is also confirmed by
the limit of~(\ref{diffsol}) and~(\ref{yc}) for $m_n \rightarrow
0\,$). We thus recover the result ${\tilde \Phi} \propto A^{-2}$ given
in~(\ref{phinophi}). Inserting a constant ${\tilde Y}$
in~(\ref{normagen}) we find the radion normalization, and finally the
coupling to SM fields given in eq.~(\ref{reslowl}). The radion mass is
obtained by expansion of the bulk eq.~(\ref{eqy}) at second order in
$l\,$, using the ansatz ($Y_0$ is the constant $0-$th order solution)
\begin{equation}
Y \left( w \right) = Y_0 + l^2 y \left( w \right) \;\;,\;\;\; m^2 =
l^2 {\tilde m}^2 \,\,.
\end{equation}
This equation can be easily solved. Imposing that $y' \left( w
\right)$ vanishes at the two branes gives the radion mass
\begin{equation}
m^2 = \frac{4 \, l^2 \, u^2 \, \left(2 k + u \right) \left( {\rm e}^{2
\, k \, w_0} -1 \right)}{3 \, k \left( {\rm e}^{2 \left( 2 k + u
\right) w_0} -1\right)} \simeq \frac{4 \, l^2 \, u^2 \left( 2 k + u
\right) {\rm e}^{-2 \left( k + u \right) w_0}}{3 \, k} \,\,.
\label{mradsl}
\end{equation}
The function $y \left( w \right)$ modifies the radion normalization at
subleading order in $l\,$; the effect can be neglected provided $l$ is
sufficiently small.

\section{Effective potential for the radion} \label{ap3}

In this Appendix we compute the effective $4$D Lagrangian for the radion
in the model~(\ref{eq:pot1}), following the original computation
of~\cite{GW}. The $4$D Lagrangian is obtained starting from an ansatz
which {\it does not} satisfy the $5$D Einstein equations of the theory, so
it should be considered only as a first, heuristic study of the system. We
discuss it here mainly to compare it with the {\it exact} $4$D description
obtained in section~\ref{quanti}. To derive it, it is convenient to use
``rescaled'' normal coordinates
\begin{equation}
d s^2 = A \left( {\bar w} \right)^2 \eta_{\mu \nu} \, d x^\mu \, d x^\nu +
w_0 \left( t \right)^2 d {\bar w}^2 \quad,\quad\quad {\bar w} \in \left[
0, 1 \right] \,\,,
\end{equation}
so that the interbrane distance is encoded in the (time-dependent)
parameter $w_0\,$. The advantage of this parameterization is that the
kinetic term for $w_0\,$ is very easily obtained. As in~\cite{GW}, we
neglect the backreaction of the scalar field $\phi$ on the metric; hence,
the analysis is restricted to the small $l$ limit (in addition, the
effective Lagrangian only refers to the radion field, while the
result~(\ref{normagen}) is valid for all the physical modes). We
``promote'' the parameter $w_0$ to be time-dependent, so that
\begin{equation}
A = {\rm e}^{-k \, {\bar w} \, w_0 \left( t \right)} \,\,.
\label{aeff}
\end{equation}
The kinetic term for $w_0$ is obtained by integration over ${\bar w}$ of
the $R_{\left( 4 \right)}$ term in~(\ref{adm}), which gives
\begin{equation}
{\cal L}_K = M^3 \int_0^1 d {\bar w} \, w_0 A^4 \frac{6 \, \ddot{A}}{A^3}
= \dots = \frac{1}{2} \left[ \sqrt{\frac{6 \, M^3}{k}} \, {\rm e}^{- k w_0
\left( t \right)} \right]^{{\small \bullet}\,2} \equiv \frac{1}{2}
\dot{\gamma}^2 \left( t \right) \,\,,
\end{equation}
where $\gamma$ is the canonically normalized field (a further contribution
to ${\cal L}_K$ is given by the kinetic term of $\phi\,$, but it is
subdominant in $l\,$).

The effective potential for the radion is obtained by integrating the
action over ${\bar w}\,$. The O$\left( l^0 \right)$ part cancels,
corresponding to the cancellation of the $4$D cosmological constant in the
RS model. Hence, we effectively integrate the action for the scalar field,
after subtracting the bulk cosmological constant and the brane tensions
from the bulk/brane potentials
\begin{equation}
- V_{\rm eff} = 2 \int_0^1 d {\bar w} \, A^4 w_0 \left[ - \frac{\phi'^2}{2
\, w_0^2} - {\tilde V} \right] - A \left( 0 \right)^4 {\tilde U}_0 - A
\left( 1 \right)^4 {\tilde U}_1
\label{veff}
\end{equation}
(in this Appendix, prime denotes differentiation with respect to ${\bar
w}\,$). The rescaled potentials are (cf. eqs.~(\ref{eq:pot1}))
\begin{eqnarray}
{\tilde V} &=& \left( 2 \, k \, u + \, \frac{u^2}{2} \right) \phi^2 + {\rm
O } \left( \phi^4 \right) \,\,, \nonumber\\
{\tilde U}_0 &=& - 2 \, u \, l^2 M^3 - 2 \, \sqrt{2} \, l \, u \, M^{3/2}
\left( \phi \left( 0 \right) - \sqrt{2} \, l \, M^{3/2} \right) +
\frac{\mu_+}{2} \left( \phi \left( 0 \right) - \sqrt{2} \, l \, M^{3/2}
\right)^2 \,\,, \nonumber\\
{\tilde U}_1 &=& 2 \, u \, l^2 M^3 \, {\rm e}^{-2 \, u \, q_0} + 2 \,
\sqrt{2} \, l \, u \, M^{3/2} \, {\rm e}^{- u \, q_0} \, \left( \phi
\left( 1 \right) - \sqrt{2} \, l \, M^{3/2} \, {\rm e}^{- u \, q_0}
\right) + \frac{\mu_-}{2} \left( \phi \left( 1 \right) - \sqrt{2} \, l \,
M^{3/2} \, {\rm e}^{- u \, q_0} \right)^2 \,.
\end{eqnarray}
Contrary to this Appendix, in the other parts of this paper the background
solution is fixed to the static configuration determined by the bulk/brane
potentials. In the present calculation, this would amount in setting $w_0
= q_0\,$ (this can be easily seen by comparing the value of $A$ at the
second brane with the potential on that brane, as given here and in
section~\ref{sec:DeWolfe}). However, our aim here is to compute the
effective potential for arbitrary $w_0\,$, so that the bulk ansatz for $A
\left( {\bar w} \right)$ and for $\phi \left( {\bar w} \right)$ has to be
given also for $w_0 \neq q_0\,$. As we shall see, the effective potential
$V_{\rm eff}$ will be precisely minimized for $w_0 = q_0\,$, in agreement
with the exact equations of the $5$D theory.

The bulk equation for $\phi$ in the unperturbed background~(\ref{aeff}) is
\begin{equation}
\phi'' - 4 \, k \, w_0 \, \phi' - w_0^2 \left( 4 \, k \, u + u^2
\right) \phi = 0 \,\,.
\end{equation}
In addition, we have the boundary conditions at the two branes, which,
in the large $\mu_\pm$ limit, read
\begin{equation}
\phi \left( 0 \right) = \sqrt{2} \, l \, M^{3/2} \quad\quad,\quad\quad
\phi \left( 1 \right) = \sqrt{2} \, l \, M^{3/2} \, {\rm e}^{-u\,q_0} \,\,.
\label{ebc}
\end{equation}
These equations give
\begin{equation}
\frac{\phi \left( {\bar w} \right)}{\sqrt{2} \, l \, M^{3/2}} = \left[
1 - \frac{{\rm e}^{\left( w_0 - q_0 \right) u} - 1}{{\rm e}^{2 \left(
2 k + u \right) w_0} - 1} \right] \, {\rm e}^{-u w_0 {\bar w}} +
\frac{{\rm e}^{\left( w_0 - q_0 \right) u} - 1}{{\rm e}^{2 \left( 2 k
+ u \right) w_0} - 1} \, {\rm e}^{\left( 4 k + u \right) w_0 {\bar w}}
\,\,.
\label{phieff}
\end{equation}
We note that, for $w_0 = q_0\,$, the expression~(\ref{phieff}) reduces
to $\phi = \sqrt{2} \, l \, M^{3/2} \, {\rm exp } \left( - u q_0 {\bar w}
\right)\,$, in agreement with eq.~(\ref{eq:sol1}). The effective
potential for the radion is now easily computed to be
\begin{equation}
V_{\rm eff} = \frac{4 \, l^2 \, M^3 \left( 2 k + u \right) \left[ {\rm
e}^{u \left( w_0 - q_0 \right)} - 1 \right]^2}{{\rm e}^{2 \left( 2 k +
u \right) w_0 } - 1 } + {\rm O} \left( l^4 \right) \,\,.
\end{equation}
which is indeed minimized (and vanishes) for $w_0 = q_0\,$.

\begin{figure}[h!]
\includegraphics[width=10cm,angle=-90]{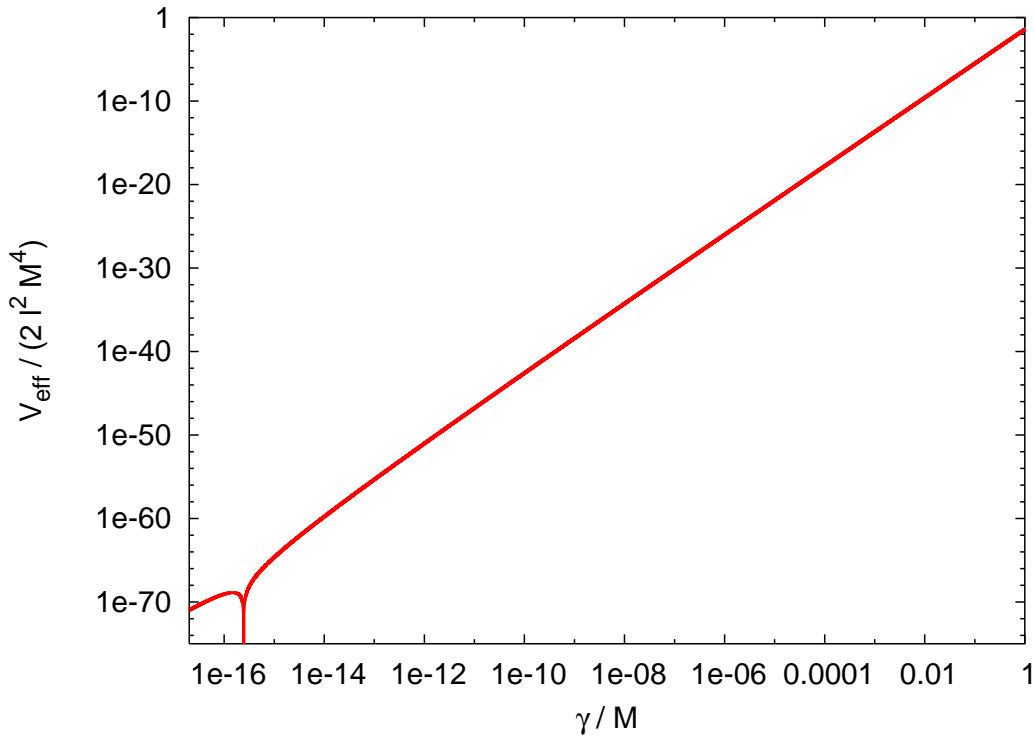}
\caption{Effective $4$D potential as a function of the canonically
normalized field $\gamma\,$. We fixed $k = M ,\, u = M/40\,$. The
parameter $q_0$ is chosen so that ${\rm exp } \left( -k\,q_0 \right) =
10^{-16}$ sets the electroweak scale at the observable brane. Note
that the interbrane distance decreases as $\gamma$ increases.}
\label{fig:veff}
\end{figure}

To summarize the computation, the effective $4$D Lagrangian for the
radion is, in terms of the canonically normalized field $\gamma\,$ and
at leading order in $l\,$,
\begin{equation}
{\cal L}_{\rm eff} = \frac{1}{2} \, \dot{\gamma}^2 - V_{\rm eff}
\quad\quad,\quad\quad V_{\rm eff} = 4 \, l^2 \, M^3 \left( 2 k + u
\right) \frac{\left[ {\rm e}^{-u \, q_0} \left(
\sqrt{\frac{k}{6\,M^3}} \, \gamma \right)^{-u/k} - 1 \right]^2}{
\left( \sqrt{\frac{k}{6\,M^3}} \, \gamma \right)^{-2 \left( 2 k + u
\right)/k} -1} \,\,.
\label{leff}
\end{equation}
In Figure~\ref{fig:veff} we show $V_{\rm eff} \left( \gamma \right)$
for a specific choice of the parameters compatible with the solution
of the hierarchy problem (see the main text). We note that the
interbrane distance {\it decreases} as $\gamma$ increases. The limit
$\gamma \rightarrow 0$ ($w_0 \rightarrow \infty$) corresponds to
infinite distance between the two branes, while the two branes
coincide for $\gamma = \sqrt{6 \, M^3 / k}$ ($w_0 = 0\,$). The
potential $V_{\rm eff}$ sharply increases as the two branes approach
each other, due to the strong gradient energy of $\phi\,$. On the
contrary, $V_{\rm eff}$ vanishes at very large distances (small
$\gamma\,$). In addition, $V_{\rm eff}$ has a minimum at finite
$\gamma \equiv \gamma_0\,$, corresponding to $w_0 = q_0\,$. This is
the value of the radion set by the stabilization mechanism.

We should caution against the use of this potential for a strongly
time-dependent $\gamma \left( t \right)\,$, and for $\gamma$ too far
from the minimum $\gamma_0\,$. Indeed, the starting
ansatz~(\ref{aeff}) and~(\ref{phieff}) {\it does not} satisfy the Einstein
equations of the $5$D theory even in the small $l$ limit. More
accurate answers can be obtained from the (numerical) integration of
the equations of the exact $5$D theory~\cite{bc}. Still, the effective
potential can be used to gain a qualitative understanding of the
system. We see that $\gamma$ has to be always very close to
$\gamma_0\,$, if we want the stabilization to take place. We note the
close analogy with to the problem of stabilizing the dilaton of
superstring theory from gaugino condensation~\cite{bs}: starting from
a static configuration far from $\gamma_0\,$, the two branes will move
to infinite distance, since the barrier in the potential at $\gamma
\la \gamma_0$ is too small to ``stop'' the evolution of the two
branes, unless $\gamma \simeq \gamma_0$ from the very
beginning. Although this problem is present at all $l\,$, the scale of
the potential shows that it becomes worse as $l$ decreases, since one
can start from smaller interbrane distance without $V_{\rm eff}$ to
exceed the limit of validity of the computation (e.g. $V_{\rm eff}$
smaller than the scale set by the AdS curvature).

The effective potential~(\ref{leff}) allows a very accurate
determination of the mass of the radion in the small $l$ limit
\begin{equation}
m_{\rm eff}^2 \equiv \frac{d^2 V_{\rm eff}}{d \gamma^2} \big\vert_{\gamma
= \gamma_0} = \frac{4 \, l^2 \, u^2 \, \left(2 k + u \right) \, {\rm
e}^{2 \, k \, q_0}}{3 \, k \left( {\rm e}^{2 \left( 2 k + u \right)
q_0} -1\right)} \simeq \frac{4 \, l^2 \, u^2 \left( 2 k + u \right)
{\rm e}^{-2 \left( k + u \right) q_0}}{3 \, k} \,\,,
\label{meff}
\end{equation}
where the last expression holds for strong warping. We note the
remarkable agreement with the $5$D calculation,
eq.~(\ref{mradsl}). This result contradicts some claims in the
literature of a discrepancy between the two computations for the
radion mass in the DFGK model. The existence of different answers was
reported for example in~\cite{csaki2}, end of section 6. The origin of
the discrepancy is most probably due to a miscalculation of $V_{\rm
eff}\,$ in the analysis quoted by~\cite{csaki2}: the mass~(\ref{meff})
obtained from the effective potential computed here is in excellent
agreement with the one from the $5$D equations given in~(\ref{mradsl})
and in~\cite{csaki2}.


\end{document}